\documentclass[twocolumn]{aastex631}

\newcommand\src{G1.9+0.3}
\newcommand\cha{{\sl Chandra}}

\catcode`\@=11
\newcommand{\gapprox}{\mathrel{\mathpalette\@versim>}}
\newcommand{\lapprox}{\mathrel{\mathpalette\@versim<}}
\newcommand{\propapprox}{\mathrel{\mathpalette\@versim\propto}}
\newcommand{\@versim}[2]
  {\lower3.1truept\vbox{\baselineskip0pt\lineskip0.5truept
\ialign{$\m@th#1\hfil##\hfil$\crcr#2\crcr\sim\crcr}}}
\catcode`\@=12
\newcommand{\ppy}{\% yr$^{-1}$}

\begin{document}

\title{Brightening and Fading in the Youngest Galactic Supernova Remnant G1.9+0.3: 13 years of monitoring with the {\sl Chandra} X-ray Observatory}

\author[0000-0002-5365-5444]{Stephen P. Reynolds}
\affiliation{Physics Department \\
North Carolina State University \\
PO Box 8202
Raleigh, NC 27695-8202, USA}

\author[0000-0002-2614-1106]{Kazimierz J. Borkowski}
\affiliation{Physics Department \\
North Carolina State University \\
PO Box 8202
Raleigh, NC 27695-8202, USA}

\author{Robert Petre}
\affiliation{NASA/GSFC \\
Mail Code 660
Greenbelt, MD 20771, USA}

\author{David A. Green}
\affiliation{Cavendish Laboratory \\
19 J.J.~Thomson Ave.
Cambridge CB3 0HE, UK}

\begin{abstract}

We report 13 years of Chandra monitoring of the youngest Galactic
supernova remnant G1.9+0.3, the only remnant known to be increasing in
brightness.  We confirm the spatially-integrated flux increase rate of
$(1.2 \pm 0.2)\%$ yr$^{-1}$ (1 -- 7 keV), but find large
spatial variations, from $-3$\% yr$^{-1}$ to 
$+7$\% yr$^{-1}$, over length scales as small as $10''$ or smaller.  We
observe relatively little change in spectral slope, though one region
shows significant hardening as it brightens by 1\% yr$^{-1}$.  Such
rates of change can be accommodated by any of several explanations,
including steady blast-wave evolution, expansion or compression of
discrete plasma blobs, magnetic turbulence, or variations in
magnetic-field aspect angle.  Our results do not constrain the mean
magnetic-field strength, but a self-consistent picture can be produced
in which the maximum particle energies are limited by the remnant age
(applying both to electrons and to ions) to about 20 TeV, and the
remnant-averaged magnetic field strength is about 30 $\mu$G.  The
deceleration parameter $m$ (average shock radius varying as $t^m$) is
about 0.7, consistent with estimates from overall expansion dynamics,
and confirming an explosion date of about 1900 CE.  Shock-efficiency
factors $\epsilon_e$ and $\epsilon_B$ (fractions of shock energy in
relativistic electrons and magnetic field) are 0.003 and 0.0002 in
this picture.  However, the large range of rates of brightness change
indicates that such a global model is oversimplified.  Temporal
variations of photon index, expected to be small but measurable with
longer time baselines, can discriminate among possible models.

\end{abstract}


\section{Introduction} \label{sec:intro}

The remnant of the most recent known supernova in our Galaxy, 
\src, with an age of order 100 years \citep{carlton11}, has much to teach
us about the early evolution of supernova remnants (SNRs) and about the 
supernova events that produced them.  The acceleration of particles to
extreme energies in energetic SNR shock waves is generally accepted as the
source of Galactic cosmic rays up to some energy which may or may not
reach the ``knee" in the cosmic-ray spectrum around 3 TeV \citep[e.g.,][]{blandford87}.  SNRs are a
particularly good arena in which to examine in detail the processes of
particle acceleration, as they can be spatially resolved and the shock-wave
environments can be well characterized.  SNRs exhibit relativistic electrons through synchrotron emission from radio to hard X-ray energies and in some cases through inverse-Compton upscattering of ambient photons to gamma-ray energies.  Protons manifest themselves only through inelastic collisions with thermal gas, producing charged and neutral pions which decay to secondary electrons and positrons and to gamma rays, respectively.  Gamma-ray studies
through GeV energies (studied from space with the Fermi and Agile satellites) to TeV energies (studied from the ground with air and water Cherenkov telescopes
such as H.E.S.S., HAWC, and LHAASO) can show extremely high particle energies, but spatial resolution is generally not able to localize sites of acceleration beyond a few resolution elements across a remnant.  High spatial and spectral resolution allows detailed study of properties of at least electron acceleration to TeV energies, and the correlation of the properties of the electron distribution with those of thermal gas can test detailed theories
of acceleration, generally attributed to diffusive shock acceleration (DSA).

\begin{figure}
    \centering
    \includegraphics[width=\linewidth]{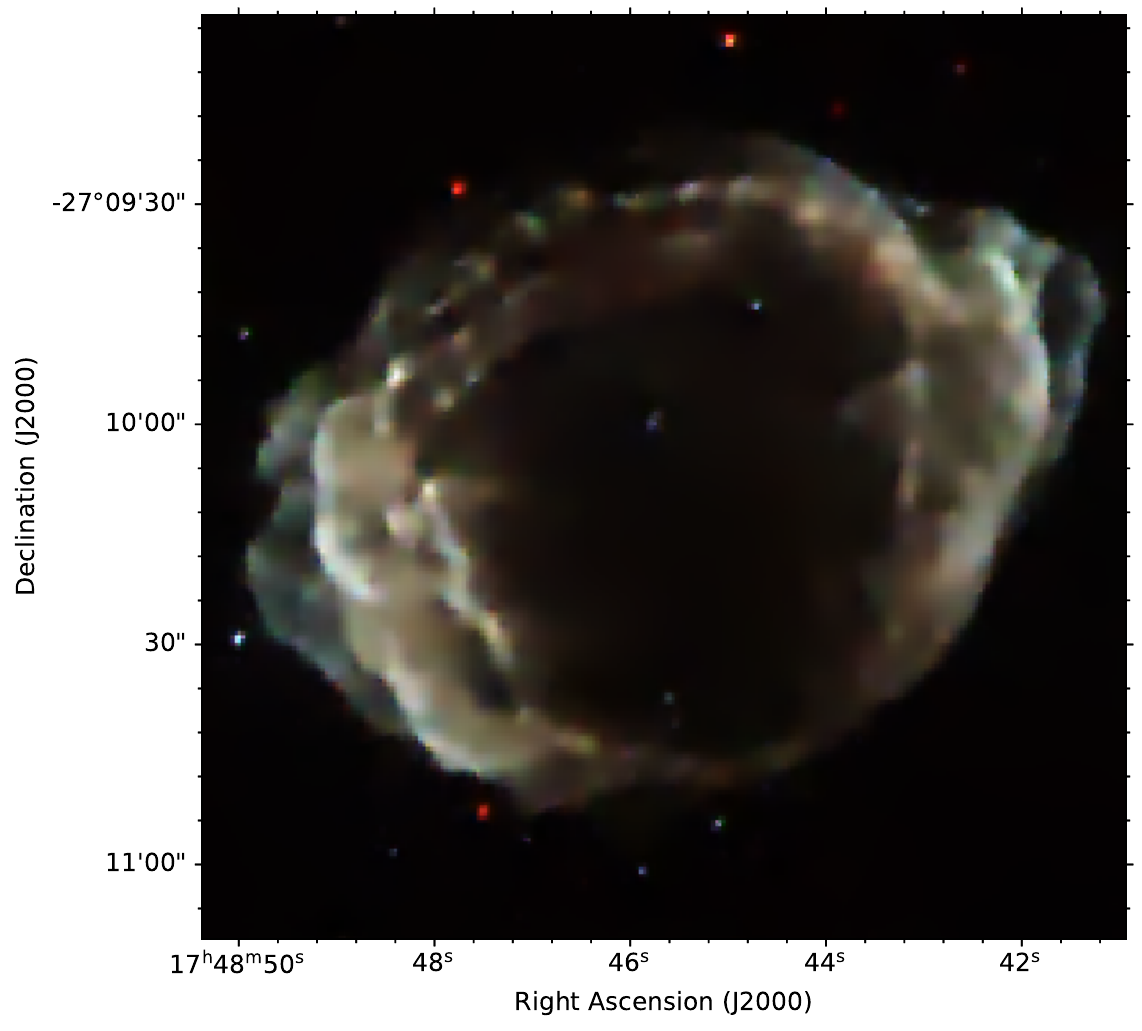}    
    \caption{2011 \cha\ image of \src\ \citep[from][]{griffeth21}. Red, 1 -- 3 keV; green,
    3 -- 4.5 keV; blue, 4.5 -- 7.5 keV.}
    \label{fig:2011im}
\end{figure}

Among supernova remnants, \src\ (Figure~\ref{fig:2011im}) is unique in the Galaxy in many ways.  While it has a dozen or so fellow remnants whose X-ray spectra are dominated by synchrotron emission, it is the only Galactic shell SNR increasing in brightness in radio \citep{green08,murphy08,luken20} and X-rays \citep{carlton11,borkowski17}.   While the integrated X-ray spectrum is featureless, small
regions have been identified showing line emission from Si, S, Ar, Ca, and Fe, 
and exhibiting line widths of order 14,000 km s$^{-1}$ \citep{borkowski13b}.  For an assumed distance of 8.5 kpc, measured expansion proper motions in radio \citep{green08,luken20} and X-rays \citep{carlton11,borkowski17} give consistent
speeds, supporting this estimate of the distance, which we shall use here. These shock velocities are the highest ever measured for a SNR, and are high even for most supernovae.  In particular, they support the identification of \src\ as the remnant of a Type Ia supernova, an identification also supported by its Galactic location and morphology \citep{reynolds08b}, though an origin in an unusual core-collapse event cannot be ruled out. The high shock velocities could result in 
gamma-ray emission as well as nonthermal X-rays, but the distance and small size
of \src\ result in low predicted fluxes, consistent with nondetections at GeV and TeV
energies \citep{brose19, hess2014}.

The general principles of particle acceleration and magnetic-field amplification 
in strong SNR shocks are now widely accepted \citep[see reviews such as][]{reynolds08a,vink12}: electron acceleration occurs by DSA, though many 
details remain cloudy; magnetic fields are amplified above adiabatic compression,
probably through nonlinear effects; and the maximum energies to which particles can reach are set by finite remnant age (or size) or radiative losses.  Abrupt changes in diffusive properties of the upstream environment could in principle also limit maximum particle energies.  Only electrons are likely to be limited by radiative losses, but it is an important and open question of whether age or losses cuts off synchrotron spectra in 
young SNRs.  DSA models cannot yet predict from first principles the relation
between shock energy and energy in particles and magnetic field; a common 
assumption is that fixed fractions of shock energy wind up in each form
(``microphysical parameters" $\epsilon_e$ and $\epsilon_B$, respectively).
Observational tests of this assumption are difficult and not encouraging \citep{reynolds21}, but for want of an alternative theoretically supported
prescription, the constant-efficiencies model is generally used.  Even if
constancy of the $\epsilon$ factors turns out to be justified, their dependences
on shock parameters (e.g., shock obliquity angle $\theta_{Bn}$ between the shock
normal and magnetic field, upstream neutral fraction, even shock velocity itself)
remain to be determined.  

An important source of information on these issues is the time-development of synchrotron emission.  In several young SNRs, variability in the brightness of small-scale features in nonthermal X-rays has been documented.  \cite{uchiyama07} report the appearance and disappearance of features on scales of about $4''$ in a year or two, in the synchrotron--X-ray--dominated SNR RX J1713.7--3946 (aka G347.3$-$0.5), which they use to argue for very high magnetic fields in these features, to obtain electron acceleration or radiative-loss timescales of order a year.   Changes have also been reported in Cas A and Tycho among other SNRs in which synchrotron emission can be at least partially isolated from the dominant thermal spectrum.  In Cas A, where the X-ray continuum between 4.2 and 6 keV is partly nonthermal,  \cite{patnaude11} report changes in the spatially integrated flux in that range of (1.5 -- 2)\% yr$^{-1}$ between 2000 and 2010, accompanied by spectral steepening. \cite{ichinohe23} describe a statistical technique for extracting spatially resolved light curves from Cas A, which they apply to the image also between 4.2 and 6 keV.  Regions of greatest rates of change are scattered in small features, though it is difficult to separate thermal and nonthermal effects. Rates of change appear to be tens of percent between 2000 and 2019.  For Tycho, \cite{okuno20} report brightening of some nonthermal ``stripes" by about 70\% over 15 years, accompanied by a significant hardening of the spectrum by 0.45 in the photon index $\Gamma$.  \cite{borkowski18} describe both brightening and fading by tens of percent in various small regions of the young (less than about 1000 years) nonthermally-dominated
SNR G330.2+1.0 over 11 years.  However, not all young SNRs show such variability; SN 1006 has fairly constant X-ray emission from its nonthermally dominated northeast limb \citep{katsuda10}.  
%Variability timescales can be used to constrain particle acceleration and loss timescales, and for small-scale features, imply very strong
%($\sim 1$ mG) magnetic fields \citep{uchiyama07}.  

However, in no case is an entire Galactic SNR observed still to be brightening at radio and X-ray wavelengths -- except one, \src, the subject of this study.  Global and local variations may have different causes.  Here we report a comprehensive analysis of brightness changes over 6 epochs from 2007 to 2020, roughly 12\% of the lifetime of \src.

\label{intro}

\section{Expansion and brightening}

\src\ was originally discovered in radio in 1984 as a compact shell source, 
proposed to be a SNR, with the smallest angular size of any Galactic SNR
\citep{green84}.  Its age was discovered in 2008 when \cha\ observations showed
a size increase of about 16\% compared to the 1984 image \citep{reynolds08b},
giving an expansion age (i.e., undecelerated) of about $140 \pm 30$ yr.   Since then, 
several global mean expansion rates at radio and X-ray wavelengths have been 
published. Brightening has been reported since the discovery; in the companion radio discovery paper \citep{green08}, a rate of $~\sim 2$\% yr$^{-1}$ is quoted.
An average rate between 1988 and 2007 of $1.22^{+0.24}_{-0.16}$\% yr$^{-1}$ was reported by \cite{murphy08}, while \cite{luken20} report $1.67 \pm 0.35$\% yr$^{-1}$, but only between 2016 and 2017.  A mean X-ray brightening rate of $1.3 \pm 0.8$\% yr$^{-1}$ was obtained by \cite{borkowski17}. While these rates scatter substantially, 
they agree on free-expansion ages of 120 -- 180 yr.  Subsequent  theoretical 
studies have deduced an expansion index $m$ (defined
by shock radius $R \propto t^m$) of about 0.7, so that the true age 
$t = mR/v \sim 100$ yr \citep{carlton11,pavlovic17,griffeth21}. 

\begin{figure*}[ht!]
    \centering
    \includegraphics[width=7.0truein]{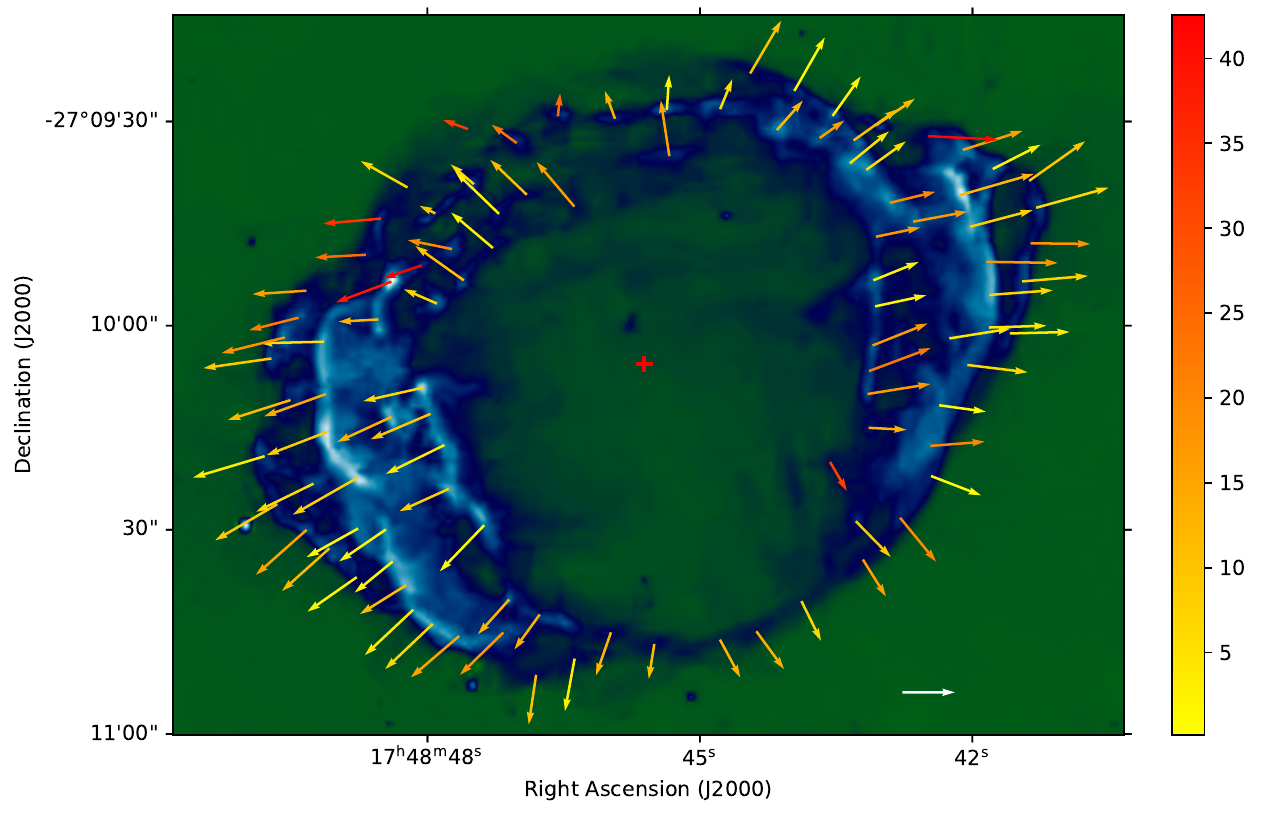}  
    \caption{Expansion between 2011 and 2020
    \citep[update of Fig.~3 in][]{borkowski17}, plotted on the epoch 2011 image. The
    white arrow indicates $0\farcs25$ yr$^{-1}$.  Arrows are color-coded
    by the angular deviation (in degrees; color bar) between the radial
    direction, with respect to the geometrical center marked with a
    red cross, and the local expansion velocity. }
    \label{fig:expan}
\end{figure*}

\begin{figure}
    \centering
    \includegraphics[width=1.0\linewidth]{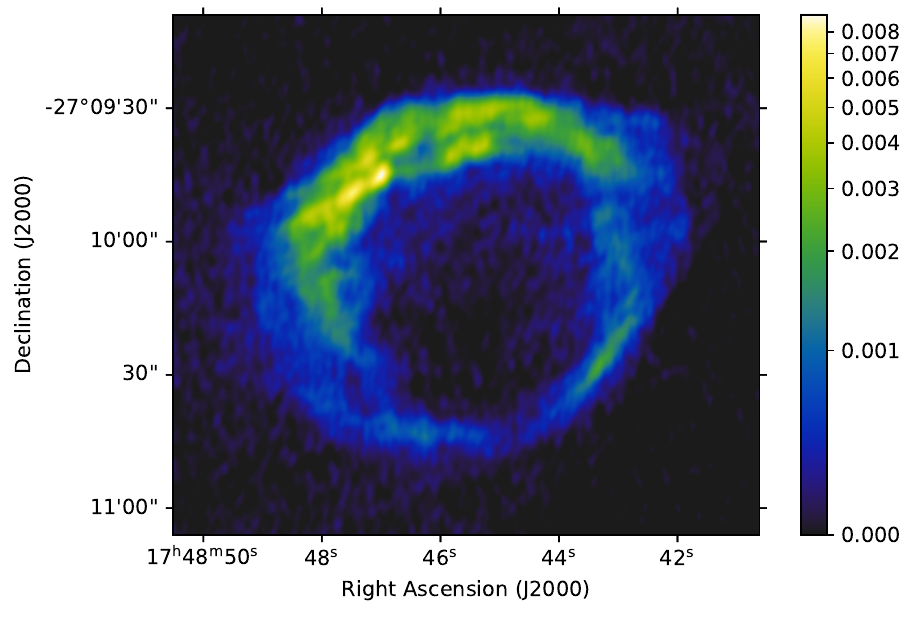}
    \caption{VLA image at 1365 MHz, epoch 2008 \citep{borkowski17}.
    The radio maximum in the NE coincides with smaller and more
    nonradial expansion rates as Figure~\ref{fig:expan} shows.
    The color scale is in Jy per $2\farcs5$ beam. }
    \label{fig:radio}
\end{figure}

\src\ has been observed by \cha\ at six epochs from 2008 until 2020, including a 1 Ms observation in 2011. This substantial time baseline allows detailed study of its evolution.  A full map of expansion motions was presented by \cite{borkowski17}, showing factors of 5 variation in expansion speed with position between 2011 and 2015, as well as significant nonradial velocity components. Figure~\ref{fig:expan} shows an update of that figure, using the same techniques to determine expansions between 2011 and 2020.  Velocities in the N and S rims are generally smaller than those in the bright E and W
limbs, but the N rim in particular shows the slowest speeds and the most
nonradial directions, confirming with smaller scatter the results of
\cite{borkowski17}.  The maximum of the radio image (Figure~\ref{fig:radio}) 
coincides with the location of some of the slowest and most nonradial motions
rather than with the maximum of the X-ray image, explained by \cite{borkowski17} as
the result of the shock encountering denser material, resulting in lower maximum
particle energies (fainter X-rays) but brighter emission at lower photon energies.
In this work we present a detailed study of the spatially resolved time evolution of \src.  

%While the integrated luminosity has been increasing fairly steadily at about 1\% yr$^{-1}$, there is a large scatter in local values, from decreases at -0.5\% yr$^{-1}$ to increases up to 7\% yr$^{-1}$.  

\section{Observations}

We have used all the observations obtained since the discovery observation in 2007 (obsIDs 6708 and 8521).  See Table~\ref{obstable}. Those newly presented here were obtained in July 2019 (obsIDs 21360 and 22275, totaling 64.8 ks) and May -- July 2020 (in chronological order 21358, 21359, 23055, 23273, 21897, 23057, 23296, 23056, 23312, and 23313, totaling 322.7 ks), for a total exposure of 387.5 ks. The rapid changes we observe in both morphology and brightness required us to analyze the 2019 data and 2020 data as separate epochs. 
All observations were performed with the ACIS-S3 chip on \cha, in Very Faint mode.

All data were reprocessed using CIAO v4.15 and CALDB v.4.10.7, then screened for particle flares. We use imaging and spectral analysis methods already described in some detail by \citet{borkowski17}. Briefly, individual pointings within each epoch were aligned using the remnant itself, while the interepoch alignment relied on numerous (30 on average) point sources. We considered only shifts in the plane of the sky, not allowing for changes in the telescope orientation.
%About 30 point sources were used on average for alignment to the 2011 epoch, with substantially fewer available only for the 2019 observation because of its short exposure time coupled with the continuing decline of the low-energy \cha\ sensitivity.
%as appropriate for the \cha\ astrometric pointing errors.
After alignment, all event files were reprojected to the reference frame of obsID 12691. Images at each epoch were extracted from merged event files created by merging reprojected intraepoch observations.

Spectra were extracted for each pointing, and at each epoch were combined together, averaging spectral and ancilliary responses. This yielded 6 spectra for each region of interest, one spectrum per epoch. We used XSPEC v.12.12.1 \citep{arnaud96} to model these unbinned spectra, assuming Poisson statistics and relying on either C-statistics \citep{cash79} or Markov chain Monte Carlo (MCMC) methods.

\label{observations}

\section{Analysis procedure}

Fluxes were obtained by modeling extracted spectra, not by counting detected photons in images, as spectral modeling properly takes into account the slow but continuous decline of \cha's sensitivity at low energies. It also allows for separation of line and continuum emission, although lines are generally weak with respect to continua in \src.

The background was modeled, not subtracted, as appropriate for spectra dominated by counting noise. We chose a background region around the remnant that did not fall into a chip gap in any of the individual pointings, and extracted background spectra at each epoch. These spectra were modeled as a mixture of sky and particle background, allowing for temporal (interepoch) variations in the particle background. We also allowed for small interepoch variations in the sky background, as this slightly improves fit quality, presumably because of the presence of systematic \cha\ calibration errors. These models were then scaled by area for each region of interest to provide us with background models.

Extracted X-ray spectra were modeled by an absorbed power law, with K-shell lines of Si, S, Ar, Ca, or Fe added if clearly present. We modeled spectra in the 1 -- 9 keV range, fitting for the column number density, the power law-index, and the logarithm of the absorbed 1 -- 7 keV continuum flux $F$.

A joint analysis of multiepoch spectra was performed to obtain best estimates
for $F$ at each epoch for each region, except for the spatially-integrated
spectrum. We found little evidence for temporal variations in spectral shapes, so we assumed time-invariant spectral shapes in our analysis for all but one region in the
southeast. All flux errors quoted in the text or shown in 
Figures~\ref{fig:sixcolor} through~\ref{fig:northwest} are
$1\sigma$ errors, while $90\%$ confidence (or credible) intervals are used for
the (normalized) flux rate $\dot F_n=\dot F/\overline{F} \equiv Q$ ($\overline{F}$
denotes estimated mean value for $F$ in the time interval from 2007 to 2020).
Note that our fits are linear, expressed in percent per year of the mean flux, 
not exponential.

Normalized flux rates and their errors were estimated by performing
weighted linear regression for 6 regions shown in Figure~\ref{fig:sixcolor}, the
biggest region in Figure~\ref{fig:rshellinnershocks}, and the spatially
integrated spectrum. An assumed systematic rms error of $2.5\%$ for $F$ was
combined in quadratures with statistical errors to estimate weights that
approximately take into account \cha\ flux calibration errors.
For the remaining regions, we performed a joint multiepoch spectral analysis in XSPEC using MCMC methods with uniform priors, assuming a linear flux increase. This assumption holds for most regions, but there are several outliers. In this method, systematic flux calibration errors are not taken into account.

A $0.5\%$ yr$^{-1}$ flux increase over one decade leads to a $5\%$ flux increase.
With a systematic rms error of $2.5\%$, \cha\ flux calibration errors
become important for rate variations less than this rough threshold. So rate
differences less than about $0.5\%$ yr$^{-1}$ are subject to poorly-understood
systematic errors, and require confirmation with future \cha\ observations.

\section{Results}
\label{results}

\begin{figure*}[ht!]
    \centering
     \includegraphics[width=4.6truein]{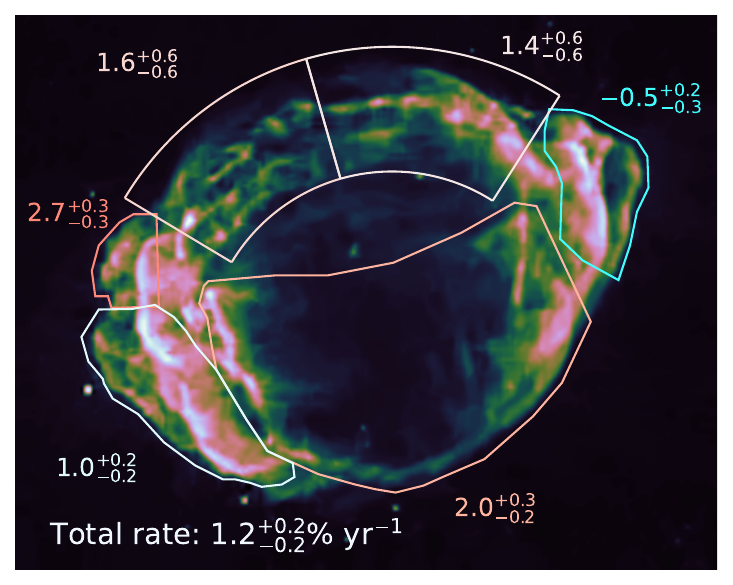}\\
     \includegraphics[width=4.6truein]{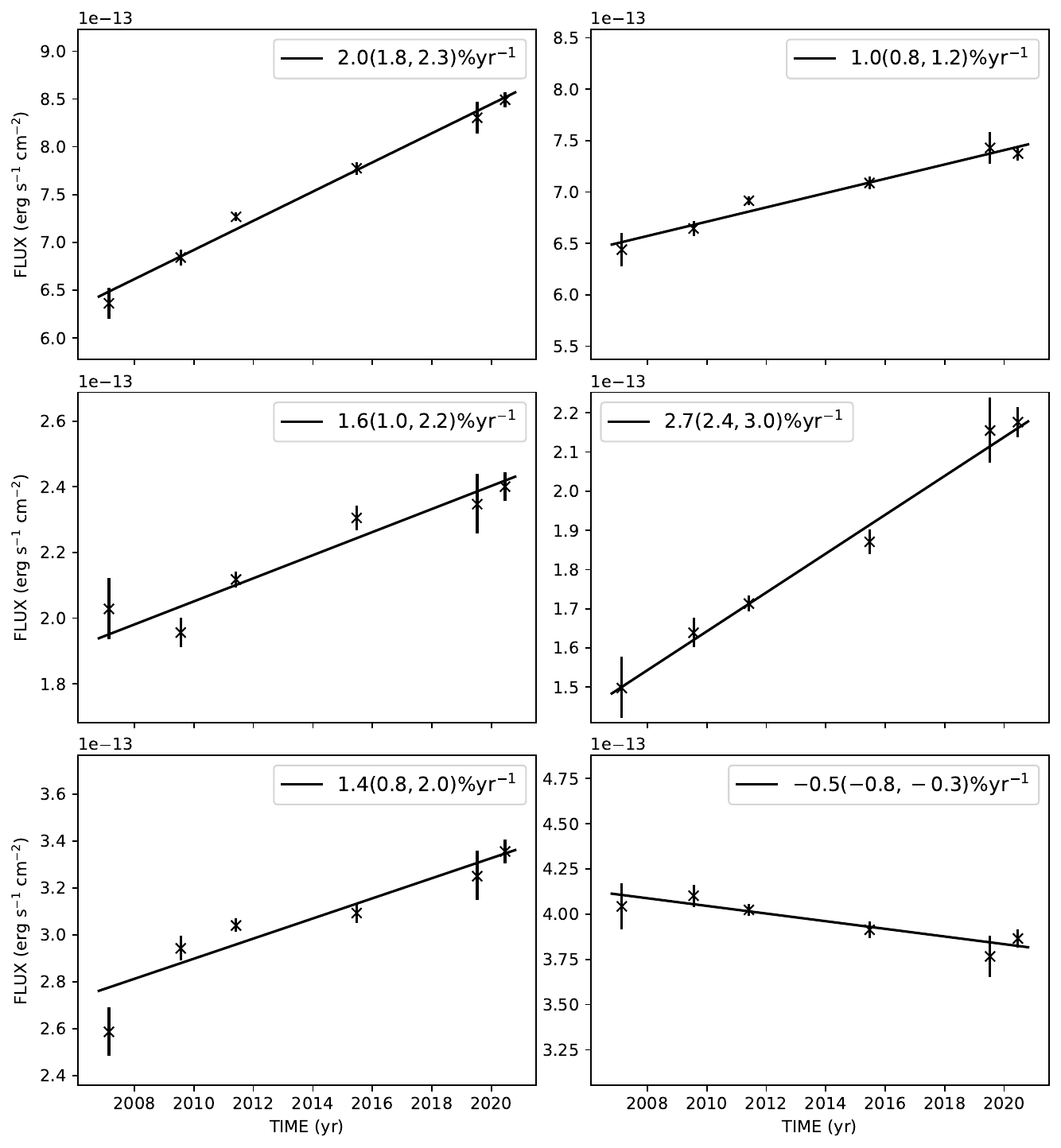}        
    \caption{Top: 2011 image with main regions indicated along with
      flux variations in percent yr$^{-1}$. Intensities are shown with the
      cubehelix color scheme of \citet{green11}.
    Bottom:  Light curves for the regions shown above.  Photon index
    $\Gamma$ (photon flux $\Phi(E) \propto E^{-\Gamma}$ ph s$^{-1}$ cm$^{-2}$
    keV$^{-1}$) for the three regions in the left column was about 2.6;
    for the right column, 2.3.}
    \label{fig:sixcolor}
\end{figure*}

\begin{figure}
    \centering
     \includegraphics[width=\linewidth]{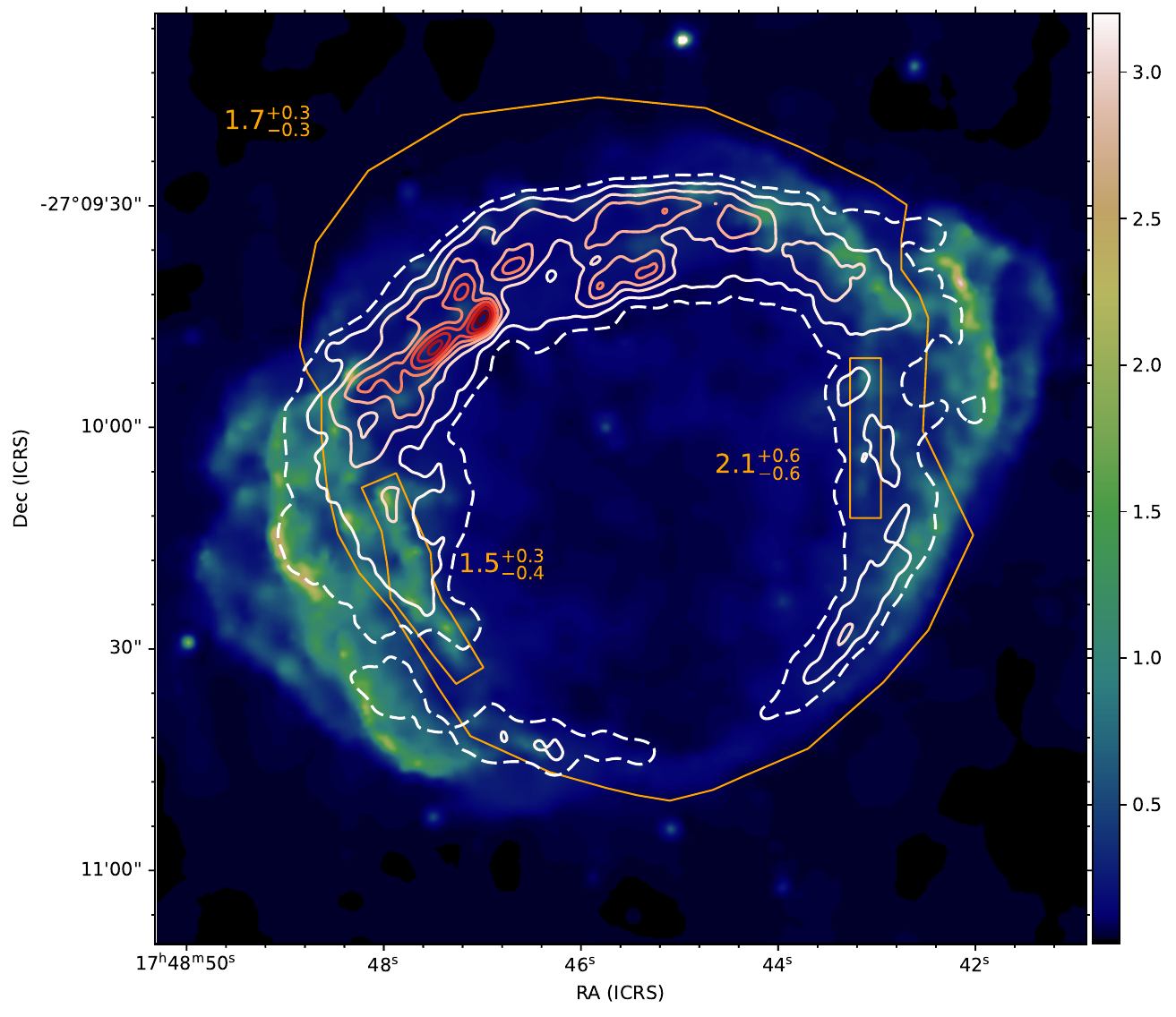}
     \caption{2009 image of \src, with selected radio contours in white and red
      (solid lines from 1 to 8 mJy beam$^{-1}$ spaced by 1 mJy beam$^{-1}$, and
      dashed line at 0.5 mJy beam$^{-1}$).  The outer orange contour outlines
      the ``radio shell" region, and two orange subregions in its interior
      indicate regions we suspect are inner shock waves. The scale is in
      counts per $0 \farcs 246 \times 0 \farcs 246$ image pixel (half an ACIS
      pixel). }
    \label{fig:rshellinnershocks}
\end{figure}

\begin{figure}
    \centering
    \includegraphics[width=\linewidth]{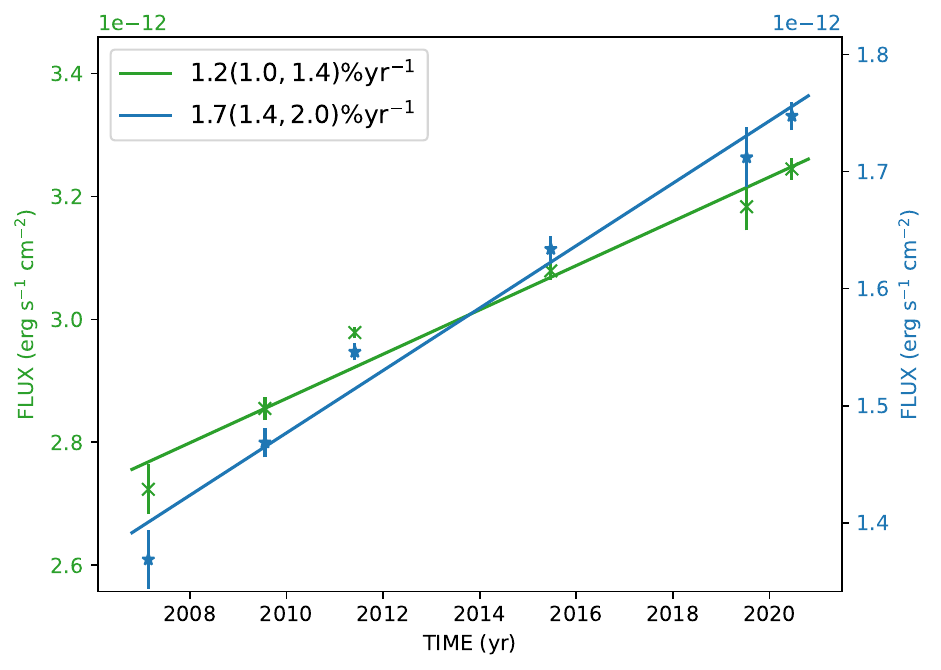}
    \includegraphics[width=\linewidth]{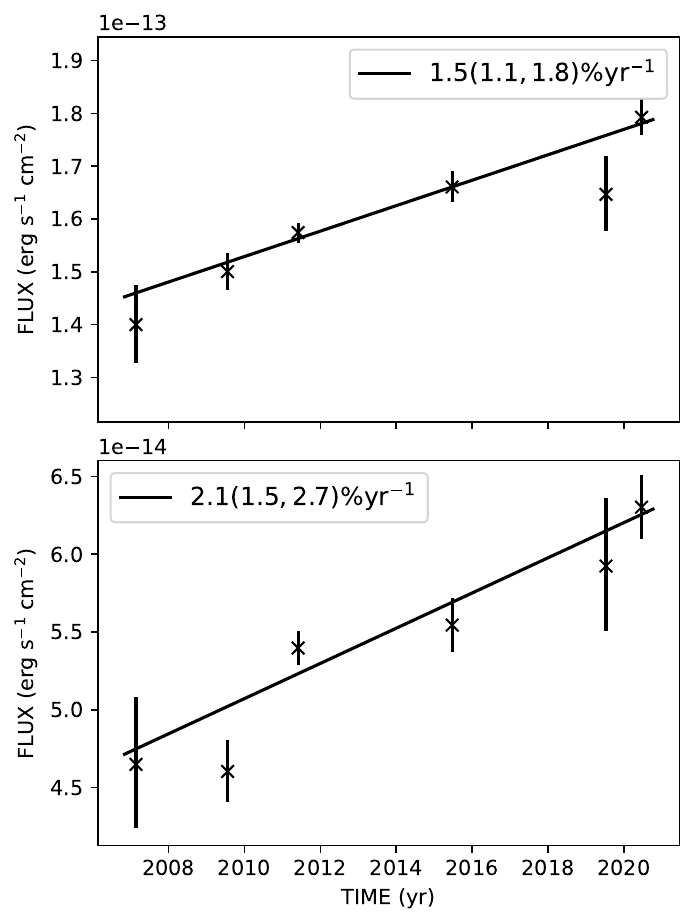}
    \caption{Top: Spatially integrated light curve for the entire remnant (green line,
    slope 1.2\% yr$^{-1}$; left $y$-axis scale) compared with that of the ``radio shell"
    region of Figure~\ref{fig:rshellinnershocks} (blue line, slope 1.7\% yr$^{-1}$); 
    right $y$-axis scale).  The ``radio shell" photon index $\Gamma$ was about 2.6.  
    Center and bottom:  Light curves for the
    two ``inner shock" regions of Figure~\ref{fig:rshellinnershocks}.}
    \label{fig:lctotrshell}
\end{figure}

\begin{figure}
    \centering    
    \includegraphics[width=\linewidth]{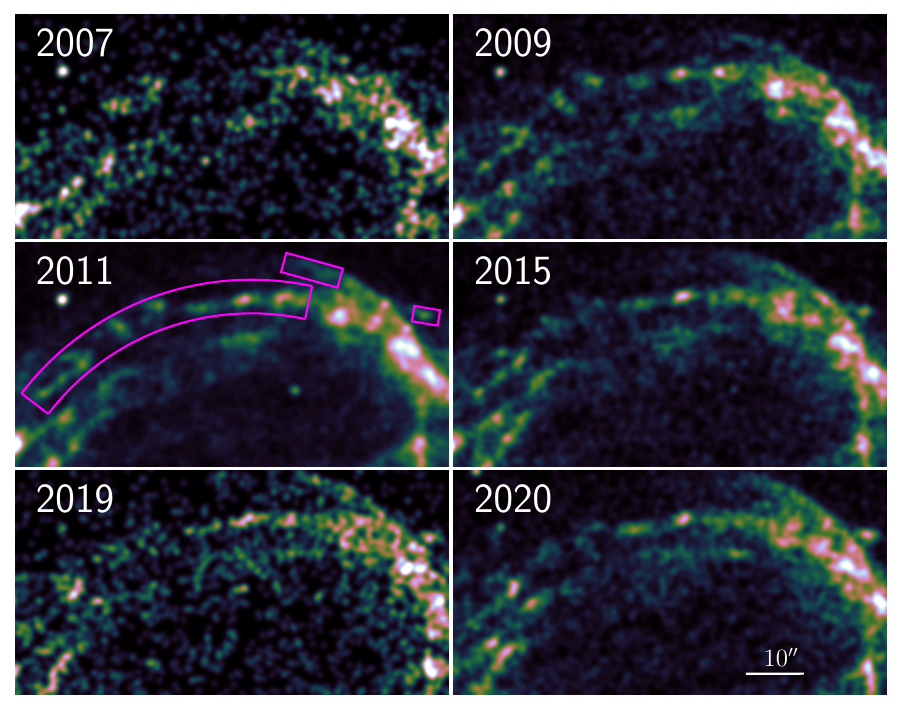}
         \caption{Brightness and morphological changes in the north rim.}
    \label{fig:northrimseq}
\end{figure}

\begin{figure}
    \centering
    \includegraphics[width=\linewidth]{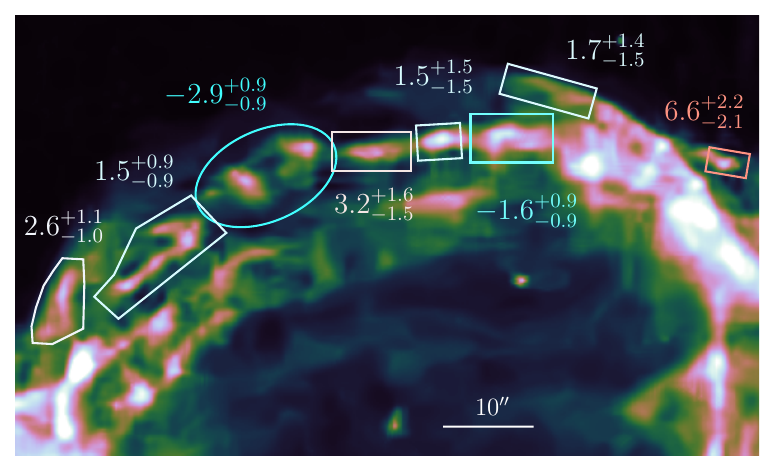}
    \includegraphics[width=\linewidth]{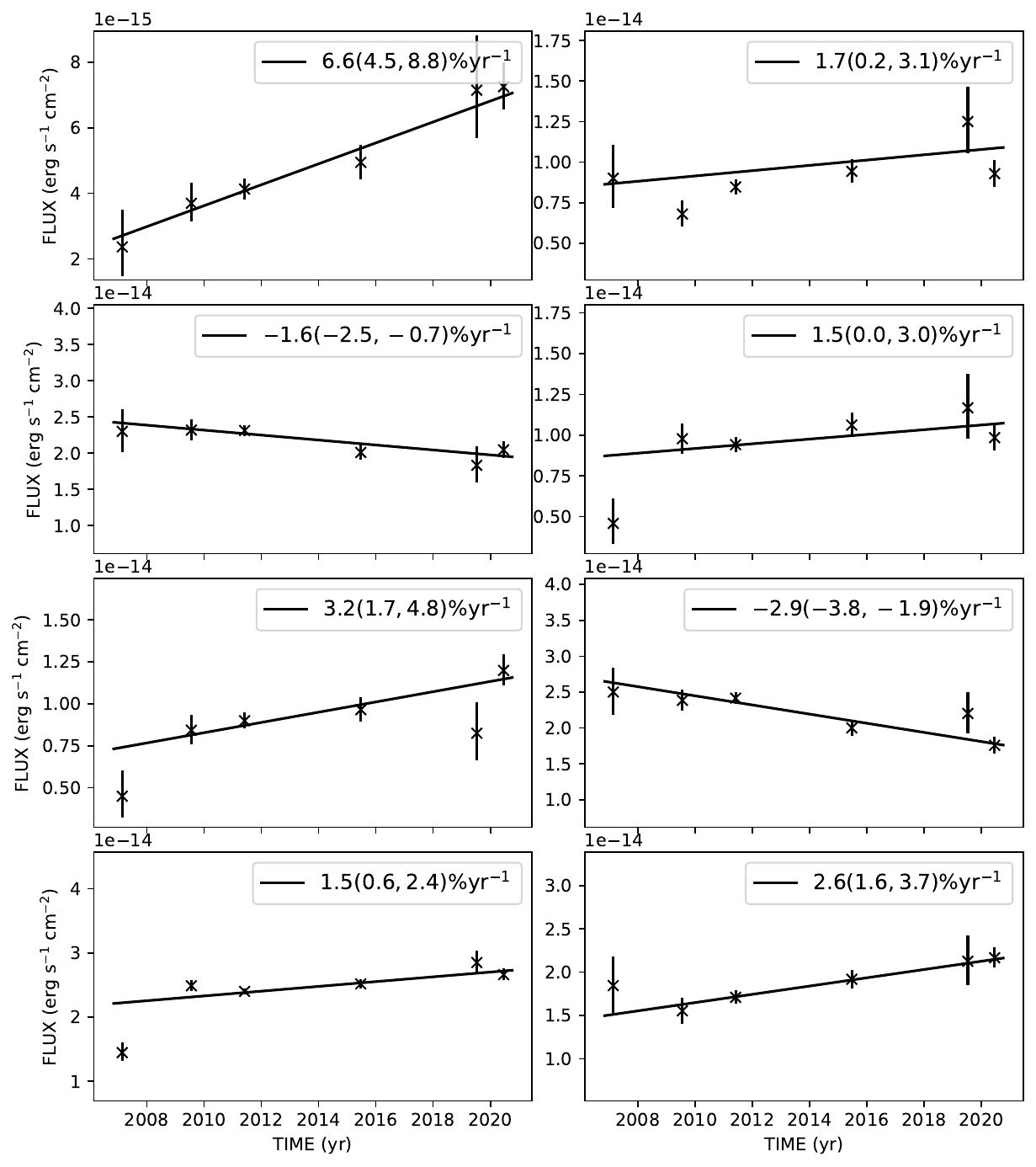}
        \caption{Top: Rates of flux change (percent yr$^{-1}$).
    Bottom:  Light curves for the corresponding regions.}
    \label{fig:northrimrates}
\end{figure}

\begin{figure}
    \centering    
    \includegraphics[width=\linewidth]{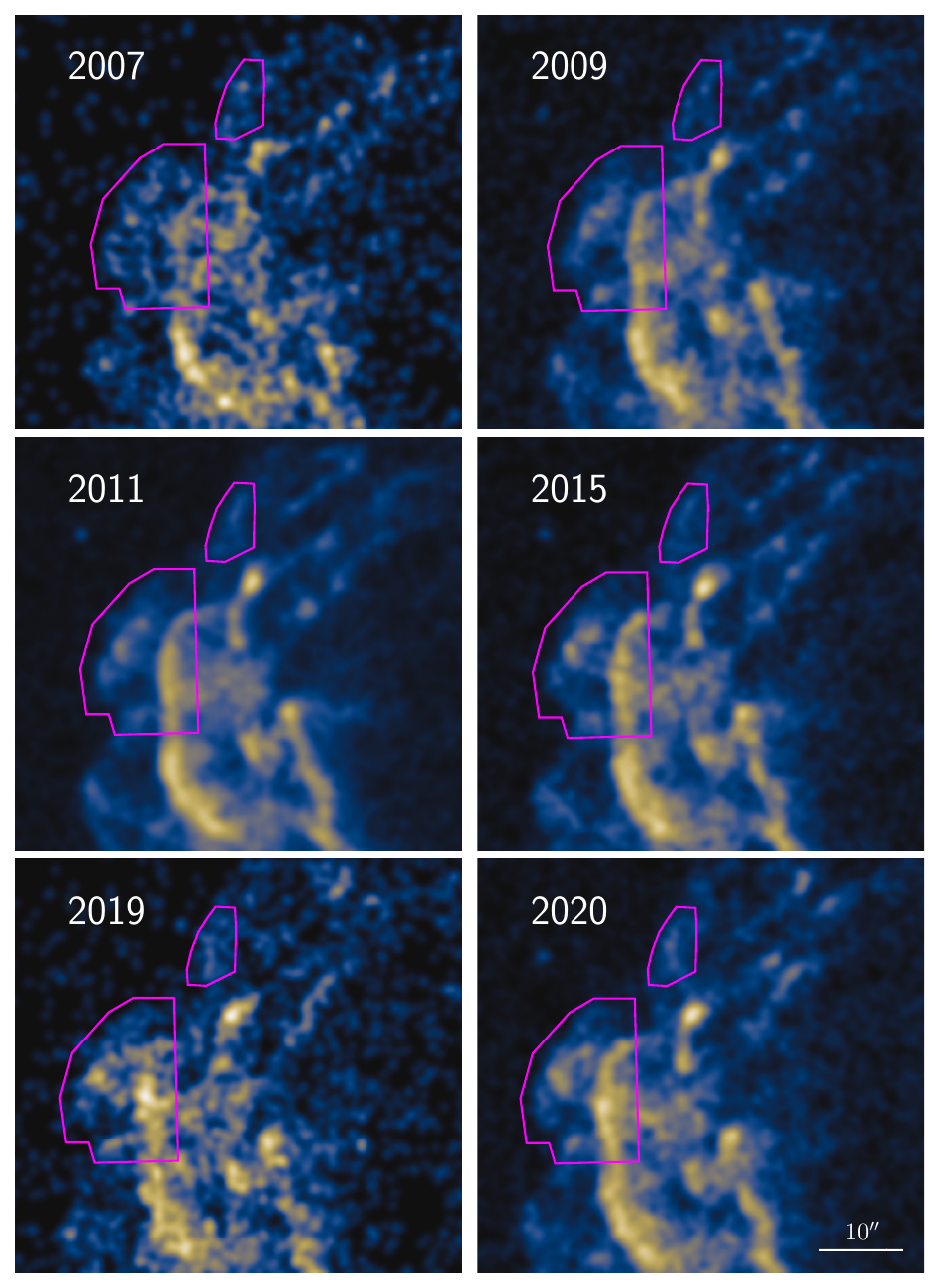}
    \caption{Changes in the northeast region of the remnant since 2007.
      Brightening rates and lightcurves are shown in Figures~\ref{fig:sixcolor}
      and~\ref{fig:northrimrates}.}
    \label{fig:varebigmodsequence}
\end{figure}

\begin{figure}
    \centering
    \includegraphics[width=\linewidth]{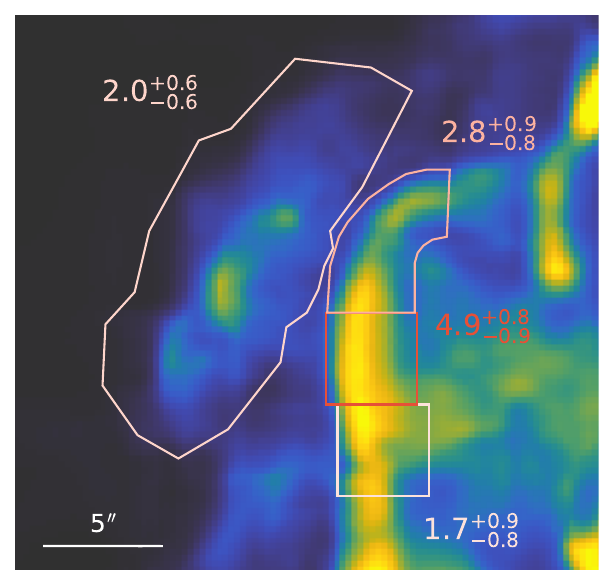}
    \includegraphics[width=\linewidth]{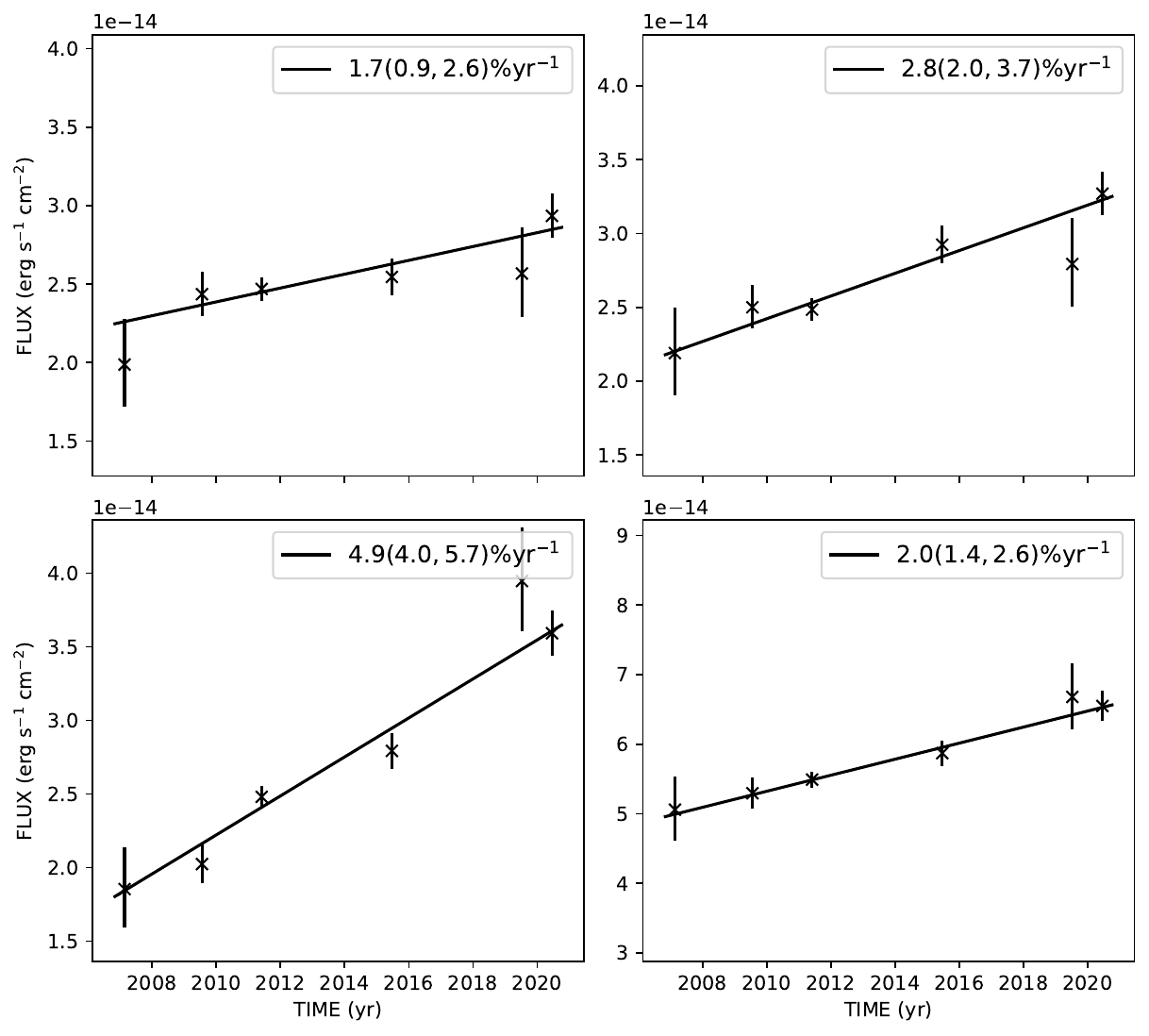}
    \caption{Top:  Rates for subregions of the area shown in
    Figure~\ref{fig:varebigmodsequence}.  Bottom: Lightcurves for those
    subregions.}
    \label{fig:var4east}
\end{figure}

\begin{figure}
    \centering
    \includegraphics[width=\linewidth]{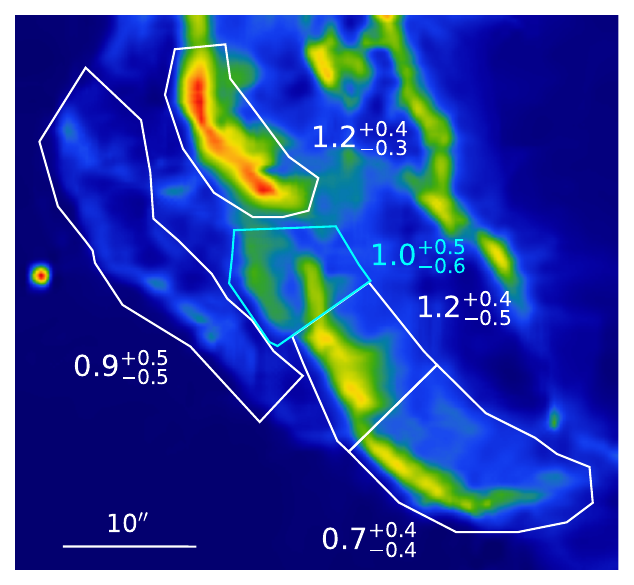}
    \includegraphics[width=\linewidth]{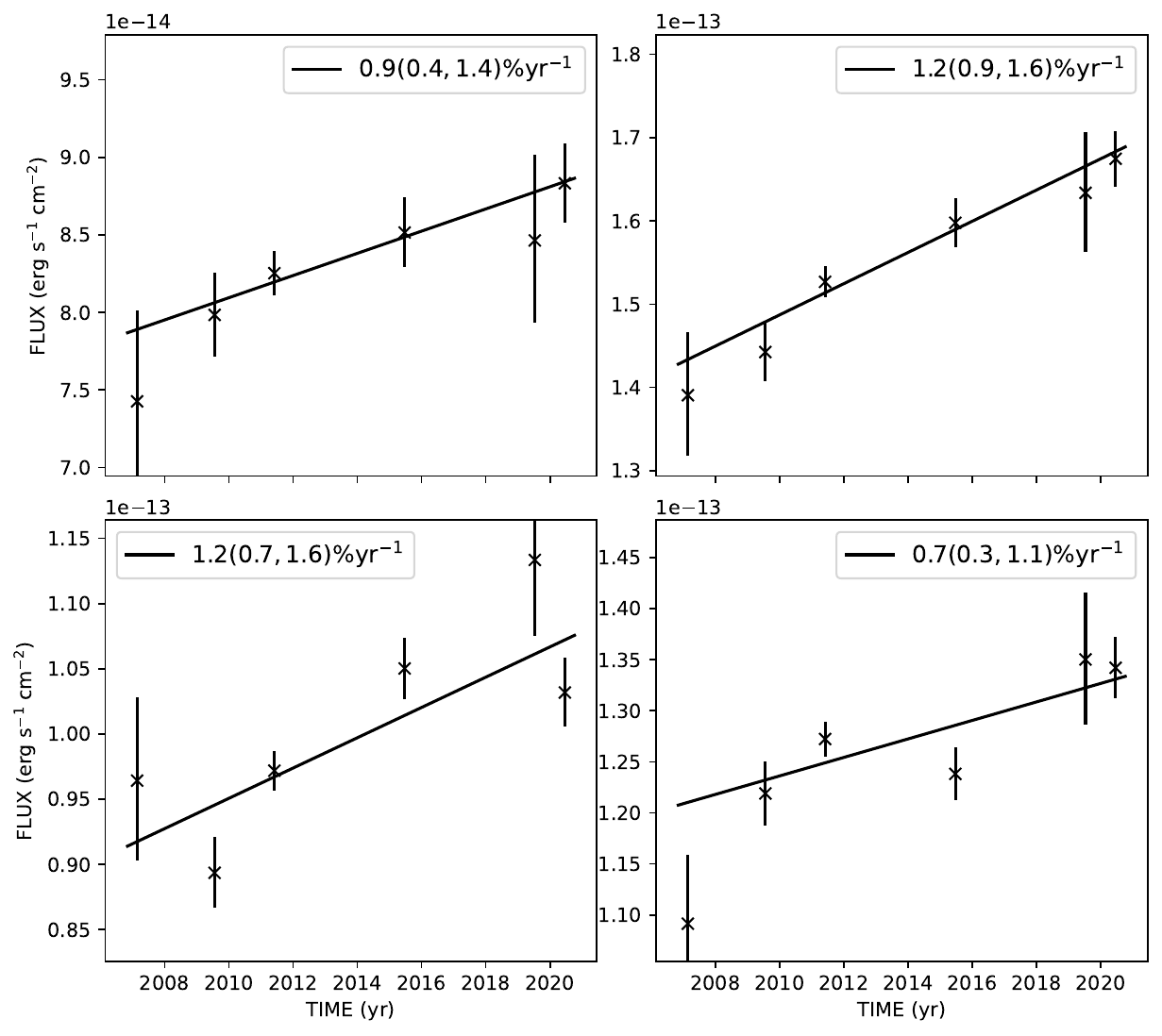}    
    \caption{Top: Changes in the southeast region.  Bottom: Lightcurves
    for the regions shown above, with the exception of the
    region outlined in blue (brightening at 1.0\% yr$^{-1}$, shown
    in Figure~\ref{fig:vargamma}).}
    \label{fig:rates5east}
\end{figure}

\begin{figure}
    \centering
     \includegraphics[width=0.8\linewidth]{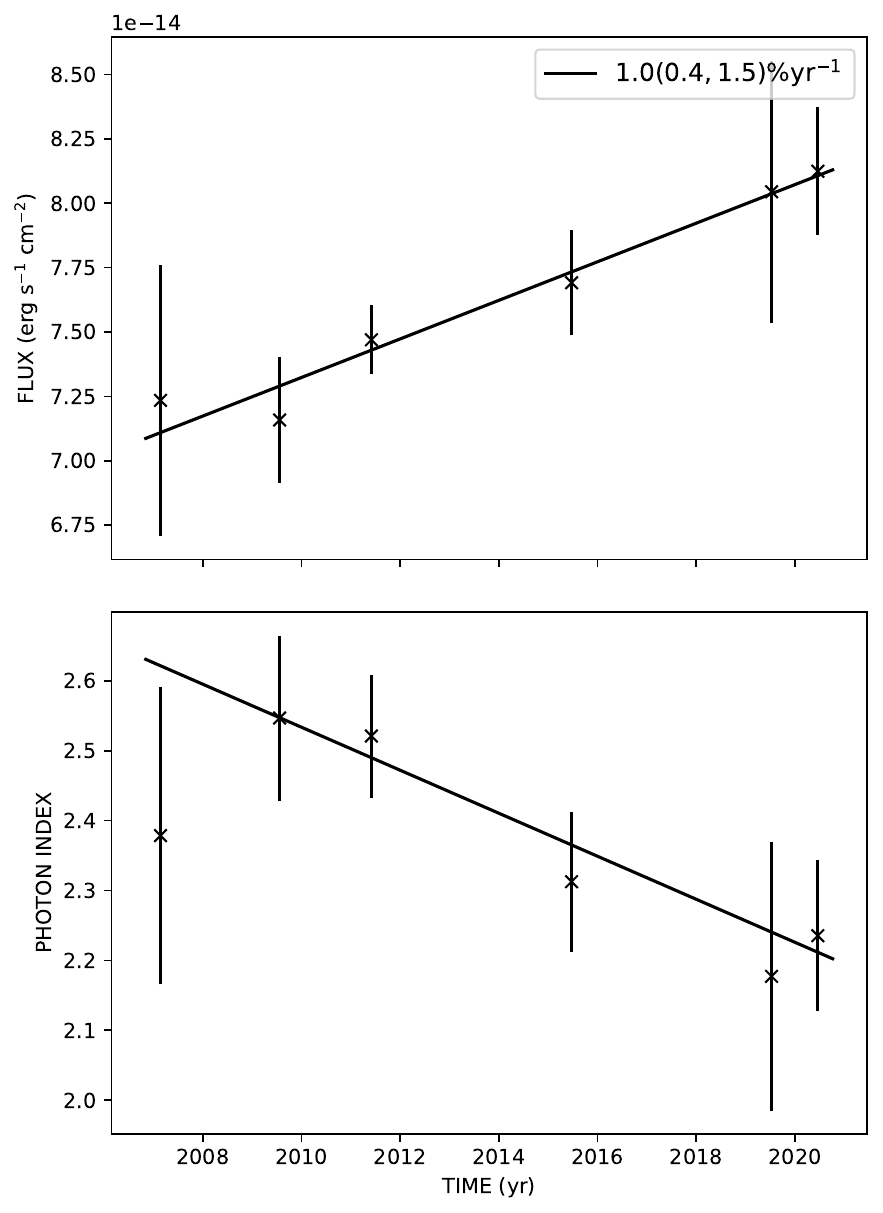}
    \caption{Light curve and variation in photon index $\Gamma$ for the region brightening
    at $1.0 (0.4, 1.5)$\% yr$^{-1}$ outlined in blue in Figure~\ref{fig:rates5east}.}
    \label{fig:vargamma}
\end{figure}

\begin{figure}
    \centering
    \includegraphics[width=\linewidth]{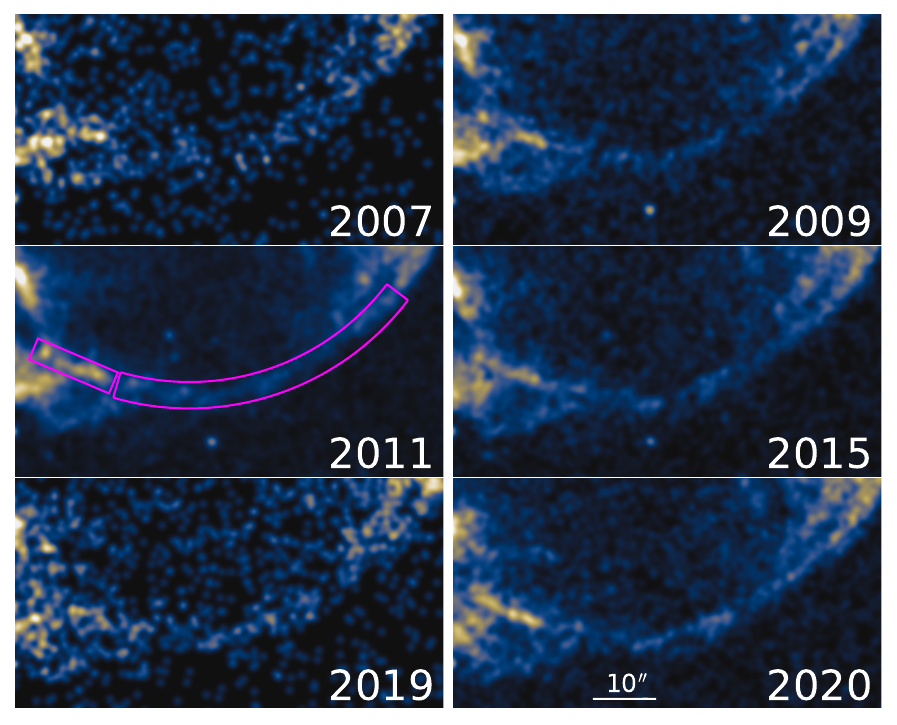}  
    \caption{Brightness and morphological changes in the south rim.}    
    \label{fig:southrimseq}
\end{figure}

\begin{figure}
    \centering
    \includegraphics[width=\linewidth]{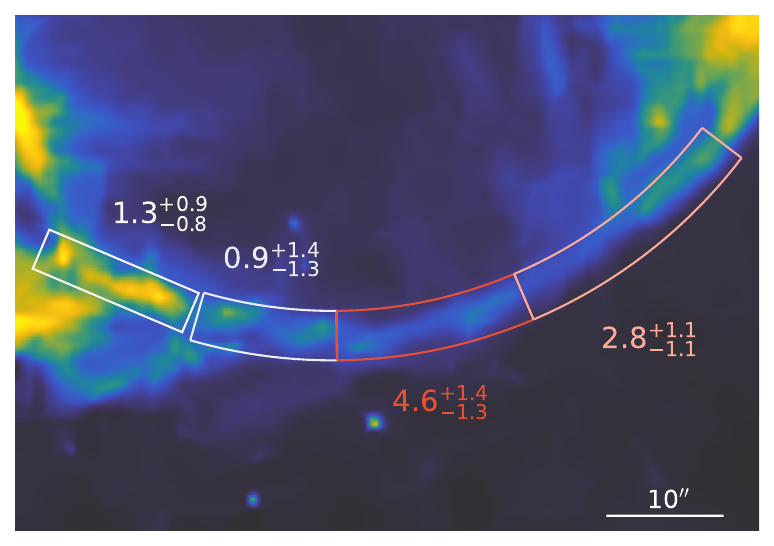}
    \includegraphics[width=\linewidth]{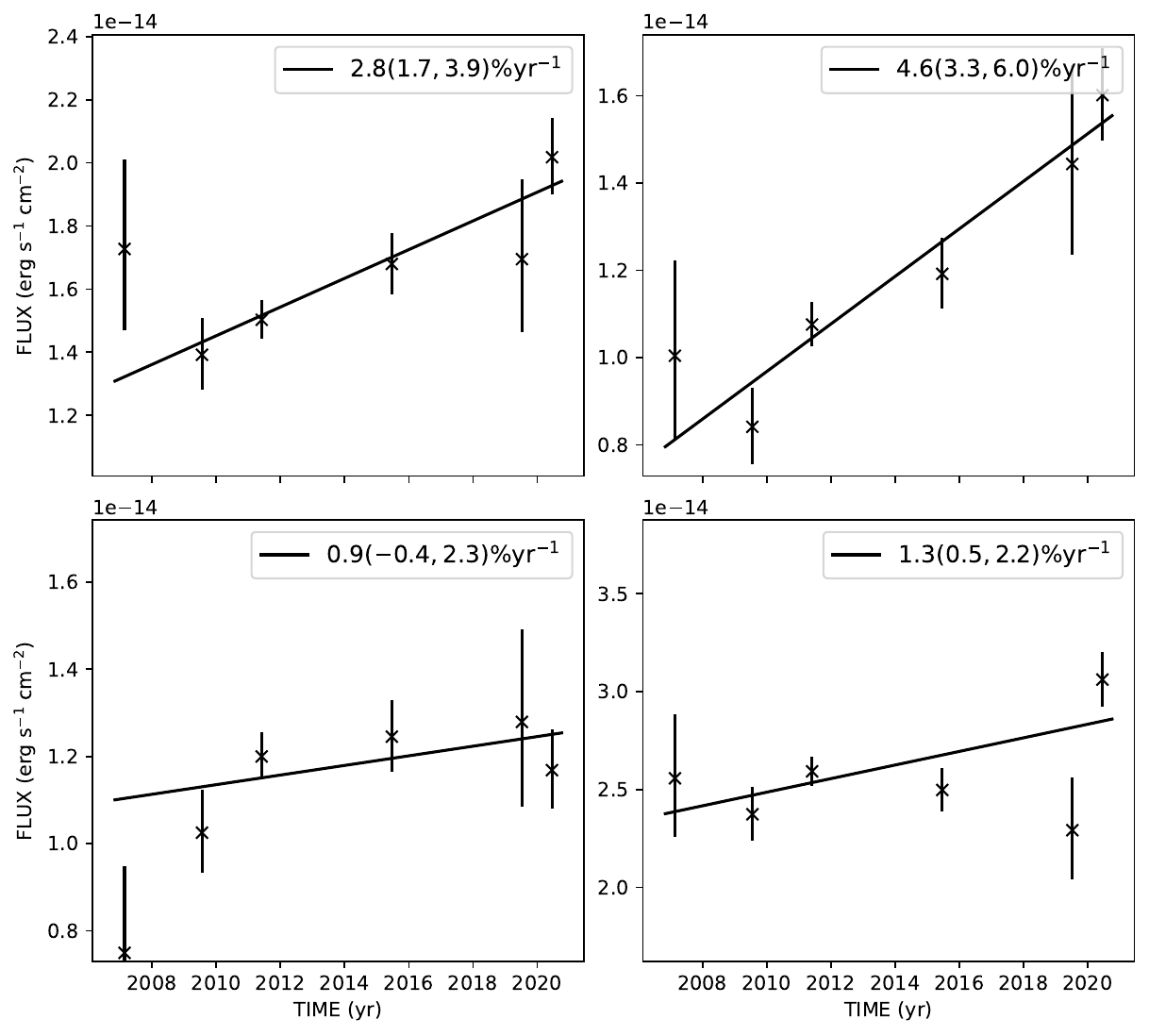}
    \caption{Top: Rates of flux change (percent yr$^{-1}$).
    Bottom:  Light curves for the corresponding regions.}
    \label{fig:southrimrates}
\end{figure}

\begin{figure}
\centering
\includegraphics[width=\linewidth]{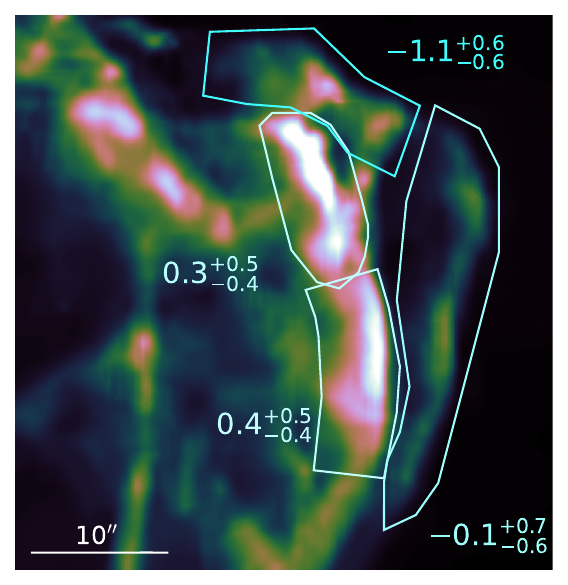}
\includegraphics[width=\linewidth]{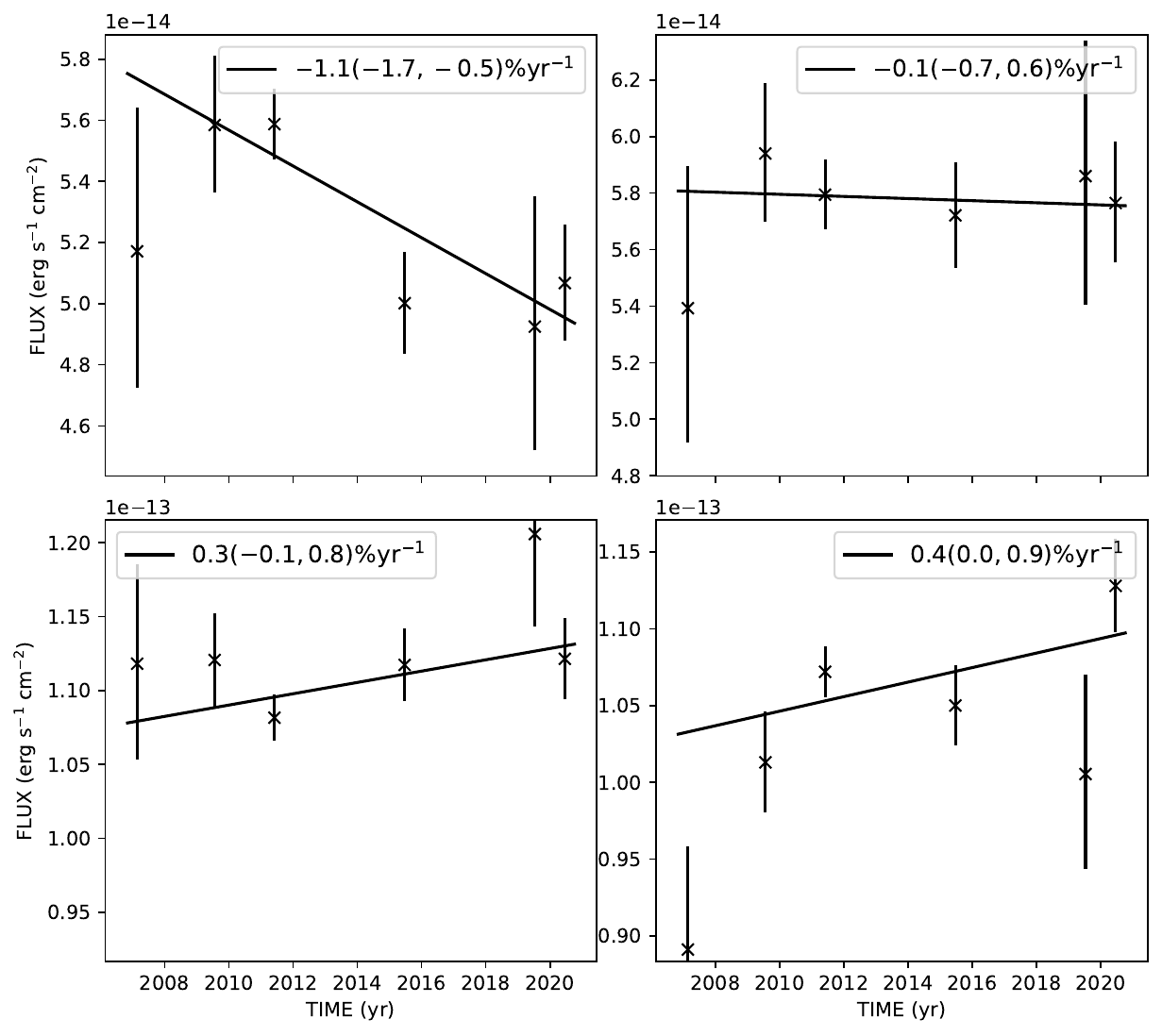}
\caption{Top:  Brightness changes in northwest regions (percent
yr$^{-1}$).
Bottom:  Corresponding light curves.}
\label{fig:northwest}
\end{figure}

Our principal observational results are displayed in Figures~\ref{fig:sixcolor} through~\ref{fig:northwest}, giving flux change rates $Q$, light curves, and
image sequences illustrating small-scale changes in morphology.
The entire remnant continues to brighten between 1 and 7 keV at a rate
consistent with that reported earlier, but with smaller errors: $(1.2 \pm 0.2)$\% yr$^{-1}$.  That is, since its age was discovered in 2007, it
has brightened by about 16\%.  However, this figure conceals a high degree
of local variation in different regions and on different length scales.
Figure~\ref{fig:sixcolor} shows rates and light curves for several large-scale portions
of \src.

%Figures

The radio and X-ray morphologies of \src\ differ markedly. 
Figure~\ref{fig:rshellinnershocks} overlays radio contours on the 2009
image.  As was shown in \cite{borkowski17}, the radio maximum in the NE
quadrant has substantially slower expansion velocities, along with highly 
nonradial motions.  That quadrant is brightening faster than the remnant average (see Figure~\ref{fig:sixcolor}).  In fact, the entire region described as ``radio shell" in Figure~\ref{fig:rshellinnershocks}
is brightening at 1.7(1.4, 2.0)\% yr$^{-1}$, faster than the remnant 
average.  The two almost linear features labeled ``inner shocks" in Figure~\ref{fig:rshellinnershocks} are also brightening
faster than average (Figure~\ref{fig:lctotrshell}).

The north rim in general is more prominent in the radio than in the X-ray image.
Figures~\ref{fig:northrimseq} and~\ref{fig:northrimrates} show the evolution of the
northernmost of three incomplete more or less parallel rims to the north (near the
edge of the radio image).  Figure~\ref{fig:northrimrates} shows the diversity of
rates, with both brightening and fading at rates between $-3$\% and $+7$\% 
yr$^{-1}$, changing sign over angular distances of $10''$ or less (about 0.4 pc).  
The global averages of Figure~\ref{fig:sixcolor} mask these strong small-scale variations.

The small regions outlined in Figure~\ref{fig:varebigmodsequence}
have some of the fastest brightening rates in the remnant at $2.7 \pm 0.3$\%
yr$^{-1}$ and $2.6(1.6, 3.7)$\% yr$^{-1}$.
On even smaller scales, Figure~\ref{fig:var4east} shows
strong spatial variations.  This region is the fastest brightening of
any of the X-ray-bright regions outside the ``radio shell" outlined
in Figure~\ref{fig:rshellinnershocks}; for the most part, the outermost
X-ray-emitting regions are brightening more slowly or fading.

The southeast region shown in Figure~\ref{fig:rates5east} exhibits changes
close to the remnant average or somewhat below it. There is convincing
evidence for spectral hardening in one location there
(Figure~\ref{fig:vargamma}). The null hypothesis, time-invariant photon index
$\Gamma$, can be rejected
with $99.9\%$ confidence using a likelihood-ratio test and a model with
linearly-decreasing $\Gamma$. We find $\dot \Gamma$ of $-0.031 \pm 0.015$
yr$^{-1}$ (errors are $90\%$ credible intervals) using a model with both $F$
and $\Gamma$ varying linearly with time.
This is a very rapid change when compared with the range of spatial variations
of $\Gamma$ seen in \src, but the errors are large.

The south rim is the least prominent in both radio and X-rays.  The morphological
correlation between radio and X-rays is strongest there. Figures~\ref{fig:southrimseq}
and~\ref{fig:southrimrates} show that morphological changes seem
less obvious; while there is no part of that rim that is fading
in X-rays, there are still strong variations along its length.

The X-ray-bright parts of \src\ in the northwest might be brightening
(Figure~\ref{fig:northwest}), but at a much slower rate than in the southeast.
No brightening is seen in the west ``ear''. Instead, the northermost region
there is fading with time, unlike its rapidly-brightening counterpart in the
northeast (both regions are distinguished by their strongly nonradial motions).

We have fit all our variations with linear brightening or fading. For the most 
part, this is generally consistent with the data, given both statistical and systematic uncertainties.  However, some regions, in particular along the south 
rim, seem to be less well described by linear trends.  A knot in the center
of the north rim (Figures~\ref{fig:northrimseq} and~\ref{fig:northrimrates}) seems
to have brightened suddenly from its 2007 value but has brightened only modestly since; similar behavior can be seen in other small features.  Some small-scale features in the north rim (Figure~\ref{fig:northrimseq}) vary rapidly, but also change morphology,
challenging comparisons between epochs.  
Continued monitoring will be necessary to document departures from linear 
changes, though such departures, in particular declines in the overall 
brightening rate, are to be expected eventually, perhaps soon.
It is also important to bear in mind that while they contribute negligibly to
the integrated spectrum, regions of thermal emission do exist and play a
more significant role in some small areas.  Furthermore, such regions are
expected to grow in strength and number as more of the supernova ejecta
are shocked.

%\subsection{Summary of results}


We summarize our observational findings as follows:

\begin{enumerate}
 
\item We find large spatial variations in flux change rates around the average
brightening of $1.2 \pm 0.2$\% yr$^{-1}$, ranging from decreases at $-$3\% yr$^{-1}$ to increases at +7\% yr$^{-1}$.

\item Radio-bright regions are brightening somewhat faster than average.  Near the
radio maximum on the NE, the brightening appears to be faster, breaking the rough
axisymmetry of the X-ray image.

\item The SE/NW extensions beyond the radio (``ears") are brightening more slowly than average in the E, and might even be fading in the W.  Those are also the regions with the fastest shocks.

\item On smaller scales ($\sim 10''$), marked variations occur, with rates in adjoining regions even differing in sign.

\item Spectra vary somewhat with position, with steeper spectra in radio-bright regions.  However, spectra do not seem to show significant variation in time, even
in regions with marked brightness changes. In one region (shown in blue in Figure~\ref{fig:rates5east}), we find $\Delta \Gamma \sim -0.4$, that is, a spectral hardening accompanying a brightening of 1\% yr$^{-1}$.

%\item What else?

\end{enumerate}

\section{Modeling brightness changes}

\subsection{General considerations}

Previous global spectral analyses of \src\ from radio to X-rays fit the data with a synchrotron model consisting of a power-law and loss-steepened tail \citep[e.g., XSPEC 
model {\tt srcut};][]{reynolds09}  or a bespoke calculation for \src\ \citep[e.g.,][]{pavlovic17}.  
Thermal emission contributes negligibly to the spatially integrated spectrum, though in small regions emission lines of intermediate-mass elements and iron are more pronounced. \citet{borkowski13b} attributed them to the shocked SN ejecta. Their heavy-element ejecta models show a negligible contribution of ejecta to the continuum emission even in small regions. We will assume all X-ray continuum emission to be synchrotron.  \cite{reynolds09} took the radio spectral index $\alpha$ ($S_\nu \propto \nu^{-\alpha}$)
to be 0.6, based on observations summarized in \cite{green19}, including
VLA observations in 2008 at both 1.45 and 4.8 GHz. Fits using the 
{\tt srcut} model in \cite{reynolds09}, anchored by a single radio flux, treated 
$\alpha$ as a free parameter, obtaining comparable values ($0.62 \pm 0.02$ in different regions).  A more recent radio study by \cite{luken20} found a considerably steeper integrated radio spectral index, $\alpha = 0.81 \pm 0.02$.  However, the flux densities at higher frequencies reported
by \cite{luken20}, from ATCA 2016 observations, are likely underestimates, as a mask was used that only included bright emission. This seems clear from
their 5 GHz integrated flux density of 0.25 Jy, less than half that obtained in \cite{green08}, and requiring a steep reversal of the 20-year radio brightening trend found by \cite{murphy08}. Hence the spectral index
they derive will be overestimated (too steep) by an unknown amount.  In any case, the extrapolation of this spectrum to 1 keV, where the spectral flux is observed to be $\sim 4 \times 10^{-12}$ erg s$^{-1}$ keV$^{-1}$ cm$^{-2}$ \citep{reynolds09}, undershoots the X-ray value by about an order of magnitude.  This would require significant concave-up curvature in the spectral-energy distribution from radio to X-rays. Such curvature is qualitatively predicted by nonlinear DSA and shown in the models of \cite{pavlovic17}, but there the effect is much smaller, and the mean radio spectral index is taken to be 0.6.  In the simple parameterizations
below, we assume a straight power-law up to the cutoff at X-ray energies; the 
appropriate spectral-index value to use for such an approximation is then
$\alpha_{rx}$, the mean index between radio and X-rays, which we shall take
to be 0.6.  This implies a (mean!) electron energy index $s \equiv 2\alpha_{rx} + 1$ of 2.2.

\subsection{A simple model for brightness and spectral changes} \label{model}

The electron population and magnetic field required for the production of synchrotron radiation can vary in many ways.  For a pure power-law distribution of electrons in energy, $N(E) = KE^{-s}$ electrons cm$^{-3}$ erg$^{-1}$ between limits
$E_l$ and $E_h$, the basic emissivity can be written
\begin{eqnarray}
    j_\nu &=& c_j(\alpha) K B^{1 + \alpha} \nu^{-\alpha}\\
    &=& c_j(\alpha) K B^{(s+1)/2} \nu^{(1-s)/2} \ \ {\rm erg\ cm}^{-3}\ {\rm s}^{-1}\ {\rm sr}^{-1}
\label{emisr}
\end{eqnarray}
where the radio spectral index $\alpha$, defined by $S_\nu \propto \nu^{-\alpha}$, is related to the electron energy
index by $s = 2\alpha + 1$, and $c_j(\alpha)$ is a constant we need not specify at
this time.  That value incorporates an average over angles $\psi$ between the line of sight and the local magnetic-field direction.

Now in all known SNRs exhibiting a power-law synchrotron component in X-rays, that component has a substantially steeper spectrum than in radio, $\alpha_x > \alpha$, where the convention is to quote a photon index $\Gamma \equiv \alpha_x + 1$, so
that the photon flux density $\Phi(E) \propto E^{-\Gamma}$ ph keV$^{-1}$ cm$^{-2}$ s$^{-1}$.  That is, the maximum energy of electrons $E_h$, averaged over the object, is below X-ray--emitting energies.  Since an electron of energy $E$ emits the peak of its synchrotron spectrum at frequency $\nu = 1.8 \times 10^{18} E^2 B$ Hz, synchrotron photons of energy $h \nu$ keV come primarily from electrons with energies $E = 23 \left(h\nu\right)^{1/2} \left(B/100\ \mu{\rm G}\right)^{-1/2}$ TeV.  The absence of any SNR with X-ray synchrotron emission on the unbroken extrapolation of its radio power-law spectrum means that peak electron energies are below this in all known cases \citep{reynolds99,hendrick01}.  Mechanisms for limiting the maximum energy of shock-accelerated electrons include radiative losses and escape, or for very young remnants, finite time for acceleration.  Expressions for these maximum energies can be found in many references
\citep[e.g.,][]{reynolds08a}: for Bohm-like diffusion (electron mean free path $\lambda = \eta r_g$, with $r_g$ the electron gyroradius and $\eta \geq 1$ the ``gyrofactor"), 
\begin{eqnarray}
    E_{\rm max}({\rm age})({\rm TeV}) &=& 0.48\, \eta^{-1}\, B_{\rm mG}\, u_8^2\, t_{\rm yr} \ \ \ \ \ {\rm and}\\
E_{\rm max} ({\rm loss})({\rm TeV}) &=& 1.6\, \eta^{-1/2}\, B_{\rm mG}^{-1/2}\, u_8.
\label{emax}
\end{eqnarray}
Here $u_8 \equiv u_{\rm sh}/10^8 \ {\rm cm\ s}^{-1}$ is the shock velocity.  In the
presence of significant shock deceleration, Equation 3 is modified.  Appendix~\ref{emaxevol}
presents a simple model (constant velocity, $m = 1$, until an abrupt transition to $m < 1$),
but the results differ only slightly from those obtained by simply using the mean shock velocity
$R/t$ in Equation 3.

For age-limited acceleration, an approximately exponential cutoff in the electron distribution $N(E)$ is expected \citep{forman83,drury91}.  \cite{drury91} reports 
that this distribution is fit to within 10\% by a power-law with exponential
cutoff with e-folding energy $E_p = 1.8 E_{\rm max}({\rm age})$. The synchrotron radiation from such a distribution can be crudely described by the delta-function approximation, in which each electron is assumed to radiate its entire power at the peak frequency.  Then one expects a slower-than-exponential ``rolloff" of the synchrotron spectrum, $j_\nu \propto \exp\left(-\sqrt{\nu/\nu_p}\right)$, where $\nu_p = 1.82 \times 10^{18} E_p^2 B$ Hz. Since the maximum energy grows with time in this case, and the magnetic field also varies with time, extra variability is introduced into the X-ray flux compared to radio.

We can then use the delta-function approximation to generalize Equation~\ref{emisr}:

\begin{equation}
  j_\nu = c_j\,K B^{(s+1)/2} \nu^{(1-s)/2}\,e^{-\sqrt{\nu/\nu_p}} \ \ {\rm erg\ cm}^{-3}\ {\rm s}^{-1}\ {\rm sr}^{-1}.
  \label{emisx}
\end{equation}
For frequencies less than a factor 
of 30 times $\nu_p$, this approximation is good to within 50\% 
compared to the numerically calculated synchrotron spectrum from an exponentially
dropping electron energy spectrum, which drops slightly more slowly; see Figure 3
in \citet{reynolds98}.

Various effects can then produce brightness changes. 
The electron population may change in energy due to volume changes (compression or expansion), changes in the electron injection or acceleration rates, or radiative energy losses.  The magnetic-field strength may change due to volume changes, magnetic-field amplification in shock waves or turbulence, or reconnection.  As a result, changes in radio and X-ray flux or brightness can be attributed to a wide range of causes.

\begin{enumerate}
    
    \item {\bf Evolutionary changes.} Ongoing acceleration of particles and
    amplification of magnetic field due to evolving shock processes will 
    change both $K$ and $B$.  Simple prescriptions can be made to describe
    these processes, although reality is likely more complicated.  Such 
    changes will affect both radio and X-ray emission.  Timescales are 
    evolutionary timescales for the shock as well as shock-acceleration
    and radiative-loss timescales for the fast particles. Observed radio flux-change rates constrain the expansion index $m$ in this picture.  We note 
    that for constant efficiencies, brightening at radio frequencies 
    requires increasing the radiating volume, unless the shock accelerates 
    somehow ($m > 1$).   
    For age-limited particle acceleration, however, as the
    maximum electron energy increases, the X-ray flux can increase
    even for constant $B$ and $K$ (i.e., for constant radio flux).  
    Such an increase would have to be accompanied by
    a hardening of the X-ray spectrum, making this proposal testable.
    
    \item {\bf Discrete emission volumes.} A distinct plasma blob of fast 
    particles and magnetic field can evolve in brightness due to adiabatic 
    compression or expansion, as well as radiative losses, without ongoing 
    replenishment of particles or magnetic field from the external medium.  
    Timescales are dynamical timescale of volume changes and perhaps radiative-loss timescales.  The range of possible change rates is limited only
    by the range of plausible local expansion or contraction rates.

    \item {\bf More complex shock interactions.} In a real object, encounters
    of the blast wave with inhomogeneous surrounding material can create
    reflected shocks, oblique shocks, and regions traversed by multiple
    shocks, with complex effects on the electron distribution.  In one 
    limiting case, \cite{melrose97} showed that a particle population 
    repeatedly shocked and subject to synchrotron losses results in a 
    distribution quite different from that produced by a single shock.
    No quantitative predictions can be made for such a wide range of
    possible hydrodynamic conditions.
    
    \item {\bf Magnetic turbulence.} Even with static relativistic-electron
    spectra, strong magnetic turbulence can cause marked brightness changes
    on small scales, with the effects increasing to higher X-ray energies,
    further on the cutting-off tail of the spectrum \citep{bykov08}. The
    changes can be fairly rapid and random.  Tens of percent variations were
    found in small regions of their simulation, on timescales of order $10^{-3}$ 
    of the Alfv\'en crossing time of the regions of size $l$: $t_{\rm cross} 
    \equiv l (B/\sqrt{4\pi\rho})^{-1}$, $\rho$ the mass density. 
    %It is not clear if these
%    results scale to different mean magnetic-field strengths.  If so, semi-quantitative estimates of $B$ could be made for regions of size $l$:
    In this explanation, mean magnetic-field strengths $B$ and 
    the dispersion around those values can in principle be constrained, 
    although no analytic predictions exist.  If this explanation is correct,
    flux-change rates have no relation to electron acceleration or energy-loss
    timescales.
    
    \item {\bf Other possibilities.} One can imagine changes in absorption
    affecting radio (free-free absorption) or X-rays (heavy-element absorption),
    changes in magnetic-field orientation affecting the synchrotron emissivity, magnetic reconnection, turbulent reacceleration of electrons, or other processes.
    Most such effects would probably require abandoning the assumption of constant
    nonthermal efficiencies.  The expression for the emissivity above (Equation~\ref{emisx}) assumes an average over the angle $\psi$ between the local magnetic field and the line of sight; if the field is macroscopically 
    highly ordered (not likely, but possible), an extra factor of $(\sin\psi)^{(s+1)/2}$ multiplies that equation.  So if that angle changes markedly, large increases
    or decreases are in principle possible.
    
\end{enumerate}

We shall focus on the first two possibilities in this work, as they are virtually guaranteed to operate. In both, radio emission will evolve due to changes in the overall electron spectrum and magnetic field, as exhibited in Equation~\ref{emisr}.  Since X-ray emission results from electrons on the loss-steepened tail of their energy distribution, it can evolve differently from radio if the characteristic break or rolloff frequency changes, as is likely.

%\begin{figure}
%    \centering
%    \includegraphics[width=\linewidth]{rx2.pdf}
%    \caption{Red: 2009 radio image.  Green: 2011 X-ray image.  The
%    bright limbs in X-rays bound the radio emission, but extend well 
%    beyond in the east and west (``ears"). }
%    \label{fig:radioXray}
%\end{figure}

\subsection{Quantitative inferences}
\label{quant}

Appendices~\ref{AppB} and~\ref{AppC} provide a set of simple calculations based 
on the emissivity of Equation~\ref{emisx}, in which flux variability due to the
first two mechanisms enumerated above can be calculated.  Before applying this 
scheme to our detailed observations, we draw a few general inferences.  To begin 
with, the overall radio brightening rate of 1\% -- 2\% yr$^{-1}$ can easily be 
accommodated by the evolutionary picture, at least in
the extremely simplistic picture of a spherically symmetric remnant encountering
uniform upstream material, a picture we know to be incorrect on several levels.
Not only is the assumption of spherical symmetry clearly incorrect at smaller
scales, but the implicit assumption of homologous expansion (uniform expansion
age, characterizing the overall expansion by a constant deceleration parameter $m$) is also at odds with observations.  \cite{borkowski14} showed that expansion ages were significantly different from the outer ``ears'' (190 yr) to the bright rim \citep[160 yr, consistent with][]{carlton11} to the inner filaments (120 yr).  For the modeling here, we disregard this additional complication. In any case, it is instructive to see what such a simple picture would imply.

Section B.2 of the Appendix (Equations~\ref{sxoft} and~\ref{defbetagamma}) shows
that the overall time-dependence of emission can be written
\begin{equation}
    S_\nu = 
    S_0 \tau^\beta\,\exp\left(-b (\tau^\gamma - 1)\right)    
\end{equation}
with
\begin{equation}
\beta \equiv m\left( \frac{s + 11}{2}\right) - \frac{s + 5}{2} \ \ {\rm and} \ \ 
\gamma \equiv \frac{5 -7m}{2}.  
%\label{defbetagamma}
\end{equation}
Here $\tau$ is the dimensionless time $t/t_0$ with $t_0$ the initial epoch when
the electron peak frequency is $\nu_{p0}$, 
and $b \equiv \sqrt{\nu/\nu_{p0}}$.  At radio observing frequencies $\nu$, 
$b \ll 1$ so $S_\nu \propto \tau^\beta$ to a high degree of accuracy.
Equation~\ref{betadef} then shows that brightening ($\beta > 0$) occurs when
$m > (s + 5)/(s + 11) = 0.55$ for $s = 2.2$.  The brightening rate $Q$ (fractional flux change/year) is given by $Q \equiv d\ln S_\nu/d\ln t/t = \beta/t$, $t$ the true
age.  Now we observe the expansion (undecelerated) age $t_{\rm exp} = t/m$; 
\cite{carlton11} report that to be $156 \pm 11$ yr.  Then $\beta = Qt = Qmt_{\rm exp}
= 6.6m - 3.6$; solving for $m$ gives values of 0.7 -- 1.0 for $s = 2.2$ and $Q$ in 
the range 0.01 -- 0.02 yr$^{-1}$.  A radio brightening rate of 2\% yr$^{-1}$ then 
demands undecelerated expansion, an unlikely possibility; a finding of a value 
close to 
2\% yr$^{-1}$ with smaller uncertainty would argue for a different brightening 
mechanism, or a departure from uniform external material -- which we already know 
to be the case.  The value $m = 1$ also implies the emitting volume increases
as $t^3$, which would surely be detectable.  The lower end of the range of $m$, 
around 0.7, is similar to the value obtained by hydrodynamic modeling \citep{carlton11}.  
That gives a true age of about 100 years.  

It is important to realize that no inference is required, or possible, on the
actual magnitude of the magnetic field, simply on its variation with time.  The 
relatively smooth brightness changes we observe in \src\ can be achieved through the systematic evolution of magnetic and electron energies.  Arguments based on equating acceleration or loss times to a few years for a fixed volume of plasma do not apply 
and need not be invoked.  
But the age of \src\ is small enough that our observation window
of 13 years is a significant fraction of it, so changes of order percent per year
are to be expected.

Additional inferences concern the X-ray emission, on the cutting-off tail of the
spectrum.  First, we have assumed that the maximum electron
energy $E_{\rm max}$ is set by the remnant age rather than by synchrotron losses.  This gives us scalings with physical parameters for $E_{\rm max}$ which will be different
for loss-limited acceleration,
for which the functional form of the spectrum is more complex as well \citep{zirakashvili07}.  

The loss-limited prescription changes the time-dependence of 
$\nu({\rm E_{\rm max}}) \equiv \nu_{\rm peak}$.  First,
from Equation~\ref{emax}, $\nu_{\rm peak}{\rm (loss)} \propto E_{\rm max}{\rm (loss)}^2 B \propto\eta^{-1} u_1^2$. Thus, only shock deceleration (or change of the diffusion 
coefficient parameterized by $\eta$) would affect the location of the high-energy
turnover in the electron spectrum: $\nu_{\rm peak}{\rm (loss)} \propto t^{2(m-1)}$, as
compared to $\nu_p{\rm (age)} \propto t^{7m-5}$. For $m \sim 0.7$, the time
exponents are $-0.6$ and $-0.1$, respectively.  That is, shock-speed variations
ought to modify the X-ray cutoff, as parameterized by the X-ray power-law slope,
much more in the case of loss-limited acceleration than for age-limited.  In fact,
the age-limited model predicts almost no evolution of the X-ray spectrum with 
time, and therefore that radio and X-ray brightening rates should be very similar.  (If the cutoff frequency grew with time, the X-ray brightening rate would be larger, and the X-ray spectral slope would also be harder, i.e., flatter.) \cite{zirakashvili07} found a considerably sharper cutoff for loss-limited acceleration; this would cause a greater variation of X-ray spectral index with change in $\nu_{\rm peak}$ than is predicted here for age-limited acceleration.

This general picture (i.e., age-limited acceleration) can
be tested for self-consistency.  Using the evolutionary model for the
integrated spectrum, as seems reasonable, Equation~\ref{xslopem} gives
the relation between observables $\alpha$ (that is, $\alpha_{rx}$) and
X-ray photon index $\Gamma$:
\begin{equation}
\Gamma = 1 + \alpha + \frac{1}{2}b\tau^\gamma
\end{equation}
with $\tau \equiv t/t_0$ as before.  
This allows us to deduce $b \equiv \sqrt{\nu_x/\nu_p}$
and hence $\nu_p$, for a fixed effective
X-ray observing frequency $\nu_x$, resulting in an observational relation
giving $B$ in terms of $\eta$, shock speed $u_s$, and age.

We first note that since $\tau$
is at most $\sim 1.1,$ the time-dependence of the term $b\tau^\gamma/2$ 
is weak:  $\gamma \equiv (5 - 7m)/2$ and $|\gamma| < 1$ for $m < 1$, as we
expect.  So $\tau^\gamma \cong 1$ and $b \cong 2(\Gamma - 1 - \alpha)$.
As described in Section~\ref{results}, values of $\Gamma$ for different regions vary from about 2.2 to 2.8.  With $\alpha_{rx} = 0.6$, we obtain $b \sim 1.2 - 2.4.$

In this approximation, $b$ is independent of the expansion parameter $m$.  We can 
apply the simple prescription of Equation~\ref{evolrate} to estimate
$m$ from observations of the fractional change rate $Q.$
%$Q \equiv \Delta S_\nu/S_\nu \Delta t \equiv \sigma/t$ where $t$ is the true
%age of the SNR.  
But we observe the expansion age $t_{\rm exp} \equiv R/v$ and $t = mR/v$.
Then Equation~\ref{evolrate} for $\sigma \equiv d \ln S_{\nu}/ d \ln t$ is an implicit equation for $m$:
\begin{eqnarray}
    \sigma &\equiv& \frac{m}{2}\left( s + 11 + 7b \right) - 
    \frac{1}{2}\left(s + 5 + 5b\right)
   = Q\,m\,t_{\rm exp} \\ &\Rightarrow&  
   m = \frac{s + 5 + 5b}{s + 11 + 7b - 2Q\,t_{\rm exp}}
   \label{Qeq}
\end{eqnarray}
where the spatially integrated brightening rate $Q$ is 1.2\% yr$^{-1}$.
With $t_{\rm exp} \cong 156$ yr \citep{carlton11}, and 
for $s = 2.2$ and $b = 2$, we find $m = 0.73$ (the dependence on $b$ is extremely weak). This value is consistent with that resulting from hydrodynamic modeling of the expansion of \src\ \citep{carlton11, griffeth21}, and gives an estimate of the
true age of \src\ of 110 yr, relative to some median of the observation interval,
near 2011, the year of the longest observation, and therefore a date of the explosion
of around 1900 CE.  In the analytic thin-shell approximation \citep{carlton11}, the mean blast-wave velocity is 18,000 km s$^{-1}$, a value consistent with that age,
taking the mean radius to be $50''$, and with the current spatially-averaged blast-wave velocity of $12,000$ km s$^{-1}$. (Somewhat higher velocities of $13,000$ km s$^{-1}$ are inferred by \citealt{borkowski17} for the ``ears" at larger radii, and are seen in line-widths of small thermally emitting regions.)

The value of $m$ we find here of 0.73 applies to a spherically symmetric model remnant
with radius $50''$, corresponding to the bright X-ray rims (and radio extent) in the E-W
direction.  For an age of 110 years (in 2011, roughly), the observed expansion ages of 190 years and 120 years for the ``ears" and the inner filaments, respectively \citep{borkowski14} then
give decelerations of $m = 0.58$ and 0.92.  If we disregard the possibility
of acceleration of the inner filaments, then $m = 1$ is an upper limit, which makes their
expansion age of 120 years an upper limit for the SNR age.  Furthermore, the 
greater deceleration (smaller $m$) for the ``ears" also suggests, from the simple model,
a slower rate of flux increase, or even, if $m \lapprox 0.55$, a decrease, compared
to the average, and a higher rate of increase for the inner regions.  These broad
predictions seem consistent with our observational results, supporting this general
picture of flux change.

From the observationally constrained values of $b$ we can find 
 $h\nu_{p0} = h\nu_x/b^2 \sim 0.5 - 0.8$ keV, taking the mean observing energy 
 $h\nu_x = 3$ keV.  A proper calculation of the age-limited maximum energy in an
 evolving shock wave $u(t)$ requires knowledge of the dynamics.  Appendix~\ref{emaxevol}
 shows a calculation based on assuming a constant expansion at $u_0$ until a transition
 time $t_{\rm tr}$ at which deceleration with $m < 1$ begins.  The result obtained
 there is very similar to that using Equation 3 with the mean expansion velocity $\langle u \rangle \cong 18,000$ km s$^{-1}$ based on the mean remnant diameter of $50''$ and an
 age of 110 years.  With 
$h\nu_{p0} = 1.82 \times 10^{18}\,E_p^2 B$
% = (1.8)^2 (1.82 \times 10^{18})\,E_{\rm max}^2 B$ 
and $E_p = 1.8 E_{\rm max}$, we find 
\begin{equation}
    B_{\mu G} = 3.5\left(\frac{h\nu_{p0}}{0.8\ {\rm keV}}\right)^{1/3}
    \left(\frac{u_8}{18}\right)^{-4/3} \eta^{2/3}
    \left(\frac{t}{110\ {\rm yr}}\right)^{-2/3}.
\end{equation}
For the fastest acceleration (Bohm limit, $\eta = 1$), this field is implausibly low, compared, for instance to a thin-rims measurement at the western edge of 320 $\mu$G \citep{reynolds21}. It is some kind of remnant average, while the thin-rim measurement applies to an unusual radio-faint part of the remnant, which turns out to be atypical in brightness evolution (the west rim seen in Figure 15 is essentially constant
in brightness at $-0.1 (-0.7, 0.6)$\% yr$^{-1}$). However, \cite{aharonian17} use the integrated \cha\ and NuSTAR data to deduce $\eta \sim 20$ to account for the surprisingly low maximum photon energy.  For that value, the above argument
gives $B \cong 26$ $\mu$G. If that average field strength fills 1/4 of the remnant volume (taking a 2 pc radius), the total magnetic energy $U_B$ is about $6.7 \times 10^{45}$ erg, and to get a flux density at 1 GHz of 1 Jy, we require a total energy in electrons (down to an assumed minimum energy $E_l = 10 m_ec^2$) of about $1.4 \times 10^{47}$ erg, so $u_e/u_B \sim 20$, comparable to ratios found in several regions of Kepler's SNR
\citep{reynolds21}. For a shock of speed 18,000 km s$^{-1}$ encountering an upstream medium with hydrogen number density $n_0 = 0.022$ cm$^{-3}$ \citep{carlton11, griffeth21}, the post-shock energy density is about $1.7 \times 10^{-7}$ erg cm$^{-3}$, and the efficiency factors are $\epsilon_e \cong 0.003$ and $\epsilon_B \cong 0.0002$.  

The low value of this mean field has important consequences for identifying the process responsible for limiting particle acceleration.  From Equations 3 and 4, the ratio of age-limited to loss-limited maximum energy is
\begin{equation}
    \frac{E_{\rm max} ({\rm age})}{E_{\rm max} ({\rm loss})} = 0.30 \eta^{-1/2} B_{\rm mG}^{3/2}\,
    u_8\, t_{\rm yr}.
\end{equation}
For $\eta = 20$, $B = 26$ $\mu$G, $u_8 = 18$, and $t_{\rm yr} = 110$, this gives
$E_{\rm max} ({\rm age}) \sim 0.5\, E_{\rm max}({\rm loss})$.  That is, the small remnant age limits the maximum energy of accelerated particles (as was assumed earlier) -- that is, the calculation is self-consistent.  The resulting maximum energy of 22 TeV would then apply to ions as well.  (Note that acceleration is age-limited even for $B_{\rm mG} = 0.004$ and $\eta = 1$.)  It should be stressed that this already oversimplified global model is crude in many ways, but the basic findings, that acceleration is limited by the remnant age and a picture explaining the global flux increase by the simplest evolving model is self-consistent, should be reliable.

\subsection{Local variations}

The total flux changes in 13 years we find are mostly between $-10$\% and $+30$\%.
These are sufficiently small that they can be accommodated with any of several possible explanations, none of which constrain magnetic-field strengths.  
Local change rates $Q$ vary between $-2.9$ and $+6.6$ percent yr$^{-1}$ in small regions. We might examine if that range can
be recovered invoking different shock properties in different regions.  Radii
around the circumference of \src\ vary by at least 20\%, demanding variations in
local expansion index $m$, which will cause variations in the flux variability
predicted by the evolutionary model. But Equation~\ref{Qeq} shows that for $s \sim 2$, $m$ is relatively insensitive to the rate of change $Q$ at X-ray wavelengths
($b \sim 2$).  Taking $s = 2.2$ and $b = 2$, $m = 0.63$ for $Q = 0$ -- that is, 
for that expansion parameter, effects of increasing emitting volume and decreasing
energy in field and particles just cancel.  A decrease at the highest rate we
find (Fig.~\ref{fig:northrimrates}), $Q = -2.9$\% yr$^{-1}$, requires $m \sim 0.5$.  The maximum increase rate with the evolutionary model assumptions is
for $m = 1$, giving $Q_{\rm max} = 3.2$\% yr$^{-1}$ for $b = 2,$ larger than
for almost all regions.  A few small regions have more rapid increases; $Q = 4.9$\% yr$^{-1}$ nominally demands $m = 1.4$ in this picture, i.e., an accelerating shock.  While this is not inconceivable, as acceleration will occur when a shock encounters a density drop, it is more likely that one of the other change mechanisms is responsible.

The volume-change explanation can account for any amount of flux variability, subject
only to the plausibility of the deduced expansion or contraction rates. It is unlikely
that large portions of the X-ray emitting material can be described by isolated blobs
disconnected from nearby material, but this explanation could certainly account for
small regions that appear or disappear with large values of $|Q|$.  

The deduction of the $b$ parameter is unchanged for this mechanism.  If the emitting
volume varies with time as $V \propto t^z$, Equation~\ref{Vtimerate2} gives 
\begin{equation}
    Qt = -\frac{2}{3} \left( s + b\tau^{2z/3} \right)\, z.
\end{equation}
We again argue that since $\tau \le 1.1$, as long as $z < 1$ we can ignore the time-dependence and take $Q = (-2/3)(s + b)z/m\,t_{\rm exp}$.  Here $m$ enters only in
inferring the true age.  Using $m = 0.7$, and with $s = 2.2$ and $b = 2,$
$z = -39 Q$.  For $-0.03 \le Q \le 0.066,$ $z = +1.2$ to $-2.6$, respectively:
a slower volume increase for the decreasing flux, and a more rapid volume decrease for
the increasing flux.  If the expansion or contraction involves all three dimensions,
$l \propto t^{+0.4}$ or $l \propto t^{-0.9}$, with $l$ a linear size.  The required
rate for the flux decrease seems entirely plausible.  The region with $Q = 0.066$ yr$^{-1}$ is the small knot in the NW (see Fig.~\ref{fig:northrimrates}), with a 
size of less than about $5''$ or $6 \times 10^{17}$ cm.  Its morphology appears to change
with time (Figure 7), making modeling difficult, but a region of this size,
shrinking by a factor $(123/110)^{-0.9} \cong 0.9$ of its original value in 13 yr, would be contracting at about 1500 km s$^{-1}$ -- a plausible fraction of the blast-wave speed.  Because of the morphology change, this may not be an appropriate
description of this particular knot, but the exercise demonstrates that high rates of brightening in small regions can be accommodated in the compression picture without requiring unreasonable parameters.  However, this example shrinkage corresponds to a decrease in size of one Chandra pixel per year, which may become measurable.  (But the expansion could be primarily along the line of sight, making
it undetectable.)  For $Q$ values between these extremes, for small regions for which
the morphology seems appropriate, the volume-change mechanism is plausible.

While a highly ordered magnetic field is perhaps not likely in the circumstances
of \src, such a field varying in direction by only a few degrees can easily
produce substantial flux changes, since $S_\nu \propto (\sin\psi)^{1.6}$, where
$\psi$ is the angle between the local magnetic field and the line of sight. For instance, a variation from $30^\circ$ to $35^\circ$ in 13 years would change the
synchrotron flux by over 20\%. 

We found one region showing statistically significant spectral changes (Figures~\ref{fig:rates5east} and~\ref{fig:vargamma}, with $\Delta \Gamma \sim -0.4$ in 13 years, as the flux rises at 1\% yr$^{-1}$).  While the sense of the correlation is expected, that is, spectral hardening should produce brightening in the X-ray band, the magnitude is too large to be described by our simple models. Equation~\ref{xslopem} gives the time-dependence of $\Gamma$, and $\Delta \Gamma = b\left(\tau^\gamma-1\right)/2$.  
But since $\tau$ varies only from 1 to 123/110 = 1.12, an unreasonably
large (negative) value of $\gamma \equiv (5 - 7m)/2$ is required, which cannot
be achieved in an evolutionary picture without unlikely circumstances such as a rapidly accelerating shock wave ($m \sim 2$).  A compression of a blob by a factor
of 2 in volume could accomplish this, but this seems unlikely as well, over the
short time available.  One of the other possibilities of our long list may be
responsible for these changes.  It is notable that only one region shows such
behavior, of all those we have studied.  Only continued monitoring of \src\ will
be able to cast light on this issue, and to narrow down the possible explanations.

\subsection{Summary of analysis results}

We summarize the results of the quantitative analysis:

\begin{enumerate}
    \item In our oversimplified spherical model, an expansion parameter $m \sim 0.7$ 
    suffices to explain the spatially integrated brightening rate of 1.2\ppy.  In the evolving picture, small regions obeying the same
    expansion law will have the same rate of flux change, of order percent yr$^{-1}$.  This value of $m$ is only slightly larger than the value at which the flux would remain constant, so small variations of $m$ can account for the various observed change rates.  We note that most earlier observations of X-ray flux changes in SNRs have found very much larger rates for small regions: appearing or disappearing altogether in timescales of order 1 year.  

    \item The evolving picture also predicts that more highly decelerated regions
    (smaller $m$) will have slower rates of flux increase, or decreases.  This is
    broadly consistent with our earlier expansion measurements giving $m = 0.58$ for the
    X-ray ``ears" and 0.92 for the inner filaments.   

    \item We assume electron acceleration to be age-limited, and find we can construct
     a self-consistent framework for the interpretation of our results.  For
     losses to limit the maximum energy to below that due to the finite age, that maximum energy would be unreasonably high.  Furthermore, loss-limited
     acceleration predicts larger variations of spectral shape with values of $Q$ than
     we typically observe, but errors are still large.  Continuing to test this 
     prediction is an important reason for further observations of \src, as it has
     implications for the maximum energies to which ions are accelerated.  The relatively
     low value of rolloff photon energy $h \nu_{m0} \sim 0.8$ keV we obtain suggests
     less rapid acceleration than the Bohm limit (i.e., gyrofactor $\eta \gg 1$).
     The age-limited model with $\eta = 20$ predicts a value for the remnant-averaged magnetic field of about 30 $\mu$G, implying maximum electron energies of about 20 TeV, a limit which should apply to ions as well.

    \item An expansion parameter $m \sim 0.7$ predicts little change in the 
    observed rolloff frequency of the emission, that is, little evolution of the photon index $\Gamma$, as we observe.  

    \item In the alternative picture of discrete magnetized plasma blobs expanding or contracting in response to dynamical changes, small volume changes could also account
    for the brightening or fading rates we observe. In fact, in this picture fractional
    volume changes $\Delta V/V$ are roughly $(1/3)\Delta S_\nu/S_\nu$, and expansion in 
    all three directions would give linear size changes $\Delta l/l \sim (1/3)$ times this,
    or of order 0.1 times $\Delta S_\nu/S_\nu$.  So a brightening of 1\ppy could be 
    achieved with an expansion speed of order $10^{-3} l$ or 100 km s$^{-1}$ for a feature
    of size scale $10''$. 
    
     \item A high degree of order in the magnetic field of small regions could
     result in substantial flux changes if the aspect angle $\psi$ between the magnetic
     field and the line of sight changes significantly.  Radio polarization
     observations (at high spatial resolution) could identify such regions of
     ordered field.

    \item An alternative picture of the origin of small-scale variability is the magnetic-turbulence picture of \cite{bykov08}, which found variations in X-ray synchrotron emission of order 10 percent in small regions, on timescales of order the regions' crossing times by turbulent characteristic velocities (presumably comparable to the Alfv\'en speed crossing times).  This occurred with static electron distributions; that is, particle acceleration and loss timescales played no role.  It is possible that such an explanation could account for some of the small-scale variability we observe.  
    %The requirement of 10-year timescales for significant flux changes in regions of \src\ of order $10''$ or **IS THIS CORRECT?** about $4 \times 10^{17}$ cm demands an unrealistically high Alfv\'en speed $v_A \equiv B/\sqrt{4 \pi \rho} \sim 10^9 $ cm s$^{-1}$, but smaller regions could flicker by substantial factors on such timescales.

\end{enumerate}

\section{Conclusions}

We have presented a spatially resolved analysis of X-ray flux evolution in a supernova
remnant, for the only Galactic SNR brightening at both radio and X-ray wavelengths.
\src\ is compact enough to fit on a single \cha\ ACIS-S chip, allowing unprecedented
sensitivity and angular resolution in X-rays over a time baseline of 13 years, about
1/8 of the remnant age.  The data we present here should open the way to extensive
analysis. The flux change rates can be compared with detailed expansion velocities
measured throughout the remnant in \cite{borkowski17}, with overall X-ray and radio
brightness, with whatever spatially resolved radio data are available, such as
polarization fraction and local radio spectral index, and with theoretical models
for SNR dynamics and particle acceleration.  It hardly needs emphasizing that
spherically symmetric modeling is not sufficient to address the variety of issues raised by the data; 3D MHD plus nonthermal modeling is required.  \src\ offers
a unique look at nonthermal physics in the fastest shocks of any SNR.

Our observational results are contained in Figures 4 -- 15.  The spatially
integrated rate of 1 -- 7 keV flux change between 2007 and 2020 is $(1.2 \pm 0.2)$\% yr$^{-1}$, consistent with earlier observations, but we find variations on
all spatial scales. Broadly, the part of the remnant brightest in radio, which has lower expansion velocities than average \citep{borkowski17}, has somewhat faster brightening rates than average, while the ``ears", which are barely detectable in radio and have the highest expansion velocities, have for the most part slower rates of brightening. On smaller angular scales, we find considerable variations.  Most regions are brightening, with a small region in the northwest increasing at $6.6 (4.5, 8.8)$\% yr$^{-1}$, but four regions show declines, the steepest of which is at $-2.9(-3.8, -1.9)$\% yr$^{-1}$. Morphological changes are also apparent; some regions clearly change shape significantly during this period, as documented in Figures 7, 9, and 13.

In this first overview of these data, we have been able to draw several 
important conclusions.

1.  A simple spectral model for synchrotron emission from a power-law distribution
of electrons with an exponential cutoff due to finite time for acceleration is
broadly consistent with the X-ray spectrum and yields a value for the expansion
index $m$ ($R \propto t^m$) of about 0.73, consistent with earlier dynamical modeling
and giving an explosion date of around 1900 CE.

2.  The same model with $m \sim 0.7$ predicts very little variation of X-ray photon 
index $\Gamma$ with either time or location.  This is basically observed, but the small variations may contain important clues on closer examination.

3.  Most surprisingly, this same successful model constrains the magnetic-field
strength averaged over the synchrotron-emitting volume to be unexpectedly low,
$\sim 30$ $\mu$G for a gyrofactor $\eta \sim 20$, though it is very likely that the field is much higher in some small regions.

4.  This low $B$ means that the age limitation on electron acceleration is more
stringent than radiative losses.  An age limitation should apply to all particles;
we deduce a maximum energy of about 20 TeV for ions as well as electrons.

5.  The range of flux-change rates we find in small regions can be explained
reasonably well by evolution of local shock 
conditions, compression or expansion of a discrete plasma volume, or perhaps
other alternatives.  Continued monitoring to detect departures from linear
changes, spectral evolution, or morphological changes, can distinguish
among the options.  

6.  As mentioned above, the photon index $\Gamma$ varies much less than the expansion
rates, consistent with the predictions of the simple model. The variations, between
2.2 and 2.6, are not well correlated with flux change rates, though regions with the hardest spectra, found in the southeast edge, have some of the slowest brightening rates.

We see small features appearing and disappearing, as has been seen in a few other
SNRs, so our linear fits to fluxes at different epochs may be too simple.  Continuing
monitoring of \src\ with \cha\ should be performed to follow evolution of the rates.
The simple estimates suggest that fairly soon, radiative losses on electrons should
become dominant, requiring a somewhat different analysis.  In any case, \src\ exhibits a stage of the evolution of the nonthermal physics of strong shocks unavailable in any other astrophysical context.

\begin{acknowledgments}
We gratefully acknowledge support from NASA through \cha\ General
Observer Program grant SAO GO9-20067X.

\end{acknowledgments}


\vspace{2mm}
\facilities{CXO, VLA}

This paper employs a list of Chandra datasets, obtained by the Chandra X-ray Observatory, contained in Chandra Data Collection (CDC) 
271~\dataset[doi:10.25574/cdc.271]{https://doi.org/10.25574/cdc.271}

\software{XSPEC \citep{arnaud96}; CIAO \citep{fruscione06}; Astronomy-oriented Python packages: astropy \citep{astropy22}, regions \citep{astropyregions22}, APLpy \citep{aplpy12}; General-purpose Python packages: matplotlib \citep{matplotlib07}, numpy \citep{2020NumPy-Array}; User-contributed colormaps: ColorCET\footnote{https://colorcet.com} \citep{kovesi15} 
}


%\software{astropy \citep{2013A&A...558A..33A,2018AJ....156..123A},  
%          Cloudy \citep{2013RMxAA..49..137F}, 
%          Source Extractor \citep{1996A&AS..117..393B}



\begin{deluxetable*}{lccc}
\tablecolumns{4}
\tablecaption{Observations for all epochs.}
\tablehead{
  \colhead{Epoch} & {ObsID} & Dates &{Exposure (ks)} }
\startdata
2007 & 6708  & 2/10/07 & 49.6 \\
     & 8521  & 3/3/07  \\
2009 & 10928 & 7/13/09 & 236.6 \\
     & 10112 & 7/18/09 &       \\
     & 10111 & 7/23/09 & \\
     & 10930 & 7/26/09 \\
2011 & 12691 & 5/9/11  & 980.5 \\
     & 12692 & 5/12/11 \\
     & 12690 & 5/16/11  \\
     & 12693 & 5/18/11 \\
     & 12695 & 5/20/11 \\
     & 12689 & 7/14/11 \\
     & 13407 & 7/18/11 \\
     & 13509 & 7/22/11 \\
2015 & 16947 & 5/4/15  & 392.3\\
     & 17651 & 5/5/15  \\
     & 17652 & 5/9/15 \\
     & 16949 & 5/20/15 \\
     & 16948 & 7/14/15 \\
     & 17702 & 7/15/15 \\
     & 17699 & 7/17/15 \\
     & 17663 & 7/24/15 \\
     & 17705 & 7/25/15 \\
     & 17700 & 8/31/15 \\
     & 18354 & 9/1/15 \\
2019 & 21360 & 7/08/19 & 64.8 \\
     & 22275 & 7/13/19 \\
2020 & 21358 & 5/16/20 & 322.7 \\
     & 21359 & 5/18/20 \\
     & 23055 & 6/4/20 \\
     & 23273 & 6/5/20 \\
     & 21897 & 6/6/20 \\
     & 23057 & 7/4/20 \\
     & 23296 & 7/5/20 \\
     & 23056 & 7/9/20 \\
     & 23312 & 7/15/20 \\
     & 23313 & 7/18/20 \\
\enddata
\tablecomments{Total exposures for each epoch are listed. The total observation time 
for all 38 observations is 2.047 Ms.
All observations were made using the ACIS-S3 chip.}
\label{obstable}
\end{deluxetable*}

\appendix
\renewcommand{\theequation}{\thesection.\arabic{equation}}

\section{Maximum particle energies in an evolving shock}
\label{emaxevol}

Equation 3 describes the maximum energy reached by particles
due to DSA in a shock of constant velocity for a time $t$.  
For an evolving shock with prescribed $u(t)$, a simple 
generalization can be made and
evaluated.  Here we assume $u_{\rm sh} = u_0$ for $0 < t < t_{\rm tr}$ and
$u = u_0(t/t_{\rm tr})^{m - 1}$ where $t_{\rm tr}$ is the transition time
from constant velocity ($m = 1$) to deceleration with fixed $m < 1$.  
\citep[This simple calculation was done for $m = 0.4$, i.e., Sedov evolution, 
in][]{reynolds98}. We fix $t_{\rm tr}$ based on the current mean
shock velocity and an assumed initial velocity $u_0$, as the time at which
the $m < 1$ extrapolation for $u(t)$ would have reached $u_0$:
\begin{equation}
    t_{\rm tr} =t \left( \frac{u}{u_0} \right)^{1/(m - 1)}.
\end{equation}
We take an initial expansion velocity $u_0$ of 20,000 km s$^{-1}$ as seen in high-velocity features of SN 2011fe \citep[e.g.,][]{silverman15}, while the current expansion rate \citep{borkowski17} gives $u = 12,000$ km s$^{-1}$.  Then, for the value of $m = 0.73$ found from the observed mean brightening rate in Section~\ref{quant}, the transition time from $m = 1$ to $m = 0.73$ is 17 yr.  This is of course a crude approximation to the true dynamics, but suffices
for a simple estimate. For constant microphysical parameter $\epsilon_B$, $B \propto u$ so $E_{\rm max} \propto u^3 \propto t^{(3m-3)}$.
So $B$(now) $= B(t_{\rm tr})(u/u_0)$ and
\begin{equation}
dE_{\rm max}({\rm TeV}) = \frac{0.48}{\eta}B_{\rm mG}(t_{\rm tr}) \frac{u}{u_0} u^2 dt
\end{equation}
and integrating, 
\begin{equation}
    E_{\rm max}(t) = E_{\rm max}(t_{\rm tr}) \left[1 + \frac{1}{t_{\rm tr}} \int_{t_{\rm tr}}^t 
     \left( \frac{t}{t_{\rm tr}}\right)^{(3m - 3)} dt \right] \\
\end{equation}
%Multiplying by 1.6 to convert from TeV to erg,
\begin{equation} 
E_{\rm max}(t) 
%= 
=E_{\rm max}(t_{tr})  \left[ 1 + \frac{1}{3m-2} \left\{ \left(\frac{t}{t_{\rm tr}}\right)^{3m-2} - 1 \right\} \right].
  \label{emaxnow}
      \end{equation}
where 
\begin{equation}
    E_{\rm max}(t_{tr})({\rm TeV)} = \frac{0.48}{\eta}B_{\rm mG}(t_{\rm tr}) u_0^2 t_{\rm tr}. 
\end{equation}
We can convert $B(t_{\rm tr})$ to $B({\rm now})$ by multiplying by an additional factor $u/u_0$.  For $m = 0.73$, $t_{\rm tr} = 17$ yr, $t = 110$ yr, $u = 12$ and $u_0 = 20$ (both in units of $10^8$ cm s$^{-1}$), the quantity in square brackets equals 3.28, and $E_{\rm max}$(now)$ = 2.8 \times 10^4 \eta^{-1} B_{\rm mG}$ erg.  Now the rolloff photon energy
$h \nu_{p0} = 1.82 \times 10^{15} h (1.8 E_{\rm max})^2 B_{\rm mG} = 1.9 \times 10^7\,B_{\rm mG}^3$
keV; since the local observed photon index $\Gamma$ gives $h\nu_{m0} = 0.8$ keV, we
infer $B \cong 3.5 \mu$G if $\eta = 1$ (Bohm-limit acceleration).  For larger values of $\eta$, $B \propto \eta^{2/3}$ rises.  The value $\eta \sim 20$ inferred by \cite{aharonian17}
(based on somewhat different arguments) gives $B \cong 26 \ \mu$G.

\section{Post-shock evolving emission}
\label{AppB}
We consider emission from a region fixed in the post-shock fluid,
focusing on time-dependence. We make standard 
assumptions about how a strong shock wave endows a volume element with a power-law electron distribution and with an amplified magnetic field: that is, that the downstream electron and magnetic energy densities $u_e$ and $u_B$ are proportional to the post-shock pressure $P_2$ with proportionality constants $\epsilon_e$ and $\epsilon_B$, respectively (see \cite{reynolds21} for references and the history of these assumptions).  (For ratio of specific heats $\gamma = 5/3$, the post-shock pressure $P_2 = (9/8) \rho u_{\rm sh}^2$, but conventional practice is to absorb the 9/8 into $\epsilon_e$ and $\epsilon_B$ and write $u_e = \epsilon_e\rho u_{\rm sh}^2$, etc.) We caution the reader that these assumptions are not well-tested against observations, and in a few cases where they can be tested, are clearly inadequate \citep[e.g.,][]{fb98}. However, they provide a useful general framework as long as their limitations are borne in mind.  Furthermore, the values of the epsilons may
vary spatially; we consider here the evolution of a region behind a local
shock wave, in which those values are at least constant in space and time.

We can then relate $K$ and $B$ to the local upstream density $\rho$ and shock speed $u_{\rm sh}$.  For our power-law electron energy distribution, the volume energy density $u_e$ obeys
\begin{eqnarray}
     u_e &\equiv& \int_{E_l}^{E_h} E\,N(E)\,dE = \frac{K}{s - 2}\left(E_l^{2-s} - E_h^{2-s}\right) \\
&\cong& \frac{K}{s-2}E_l^{2-s} \Rightarrow \ 
 K = (s - 2)E_l^{s-2} \epsilon_e \rho u_{\rm sh}^2
\label{ue}    
\end{eqnarray}
where we have assumed $s > 2$ and $E_h \gg E_l$.  So the electron energy density
is dominated by the lowest-energy electrons, and we can ignore any
contribution from electrons above some break energy at which the spectrum 
steepens or cuts off, as long as that energy is far above $E_l$.  Similarly, the magnetic-field strength $B$ is given by
\begin{equation}
  u_B \equiv \frac{B^2}{8\pi} = \epsilon_B \rho u_{\rm sh}^2 \ \Rightarrow \ 
    B = \left( 8 \pi \epsilon_B \right)^{1/2} \rho^{1/2} u_{\rm sh}.
    \label{uB}
\end{equation}

We can then write for the emissivity
\begin{equation}
j_\nu \propto \epsilon_e\, \epsilon_B^{(s + 1)/4}\, \rho^{(s + 5)/4}\,
  u_{\rm sh}^{(s+5)/2}\exp({-\sqrt{\nu/\nu_p}}).
\label{emiseps}
\end{equation}
Since $\nu_p \propto E_{\rm max}^2 B$ and for age-limited acceleration $E_{\rm max} \propto B\, u_8^2\, t$, we have 
$\nu_p \propto B^3\, u_{\rm sh}^4\, t^2 \propto u_{\rm sh}^7\,t^2$.

We are interested in the time-dependence of the emission, as conditions evolve.
Globally, \src\ is clearly in a pre-Sedov evolutionary phase, with mean expansion index $m$ (defined by $R_{\rm shock} \propto t^m$) around 0.7 \citep{carlton11}.
Then $u_{\rm sh} \propto t^{m-1}$, though $m$ may vary locally. 
Then $\nu_p \propto t^{7(m-1) + 2}$.  
Choosing a reference time $t_0$ at which $S_\nu(t_0) \equiv S_0$, we define a dimensionless time $\tau \equiv t/t_0$.
Then $\nu_p(t) = \nu_{p0} (t/t_0)^{7m - 5} \propto \tau^{7m - 5}$, with
$\nu_{p0} \equiv \nu_p(t_0).$ So the argument  of the exponential above varies with time: 
$\sqrt{\nu/\nu_p} \equiv b \tau^{(5 - 7m)/2} \equiv b \tau^\gamma$ (again, 
$b \equiv \sqrt{\nu/\nu_{p0}}$).  (The observing frequency $\nu$ is fixed.)  

Assuming a constant ambient density
$\rho$, then at frequencies low enough that the exponential factor is 1,
\begin{eqnarray}
j_\nu &\propto& t^{(m-1)(s + 5)/2}\ \ \  {\rm and} \\ 
S_\nu &\propto& R^3 j_\nu 
  \propto t^{3m + (m-1)(s + 5)/2}
\equiv t^\beta.
\label{betadef}
\end{eqnarray}
Here we have assumed homologous expansion, that is, our region expands proportionally
to the volume of the entire remnant. 
Including the exponential factor $\exp(-\sqrt{\nu/\nu_p(t)})$, then,
the overall time-dependence of emission % at frequencies above $\nu_p$ 
can be written
\begin{equation}
    S_\nu = 
    S_0 \tau^\beta\,\exp\left(-b (\tau^\gamma - 1)\right)
\label{sxoft}    
\end{equation}
with
\begin{equation}
\beta \equiv m\left( \frac{s + 11}{2}\right) - \frac{s + 5}{2} \ \ {\rm and} \ \ 
\gamma \equiv \frac{5 -7m}{2}.
\label{defbetagamma}
\end{equation}

We can then calculate the rate of flux change:

\begin{equation}
\frac{d \ln S_\nu}{d \ln t} = \beta - b \gamma \tau^\gamma \equiv \sigma \ \ {\rm so} \ \ S_\nu \propto t^\sigma.
\label{evolrate}
\end{equation}

We can also calculate the spectral slope at frequency $\nu$:

\begin{equation}
\frac{dS_\nu}{d\nu}= S_\nu \left[ \frac{-\alpha}{\nu} - 
  \frac{1}{2\nu} \left( \frac{\nu}{\nu_p} \right)^{1/2} \right]
  \end{equation}

and therefore

\begin{equation}
\frac{d \ln S_\nu}{d \ln \nu} = 
  -\left[ \alpha + \frac{1}{2} \left( \frac{\nu}{\nu_p} \right)^{1/2} \right].
  \label{slope}
  \end{equation}
  
\subsection{Radio emission}
\label{evolveradio}

Even for the kinds of field strengths found in young supernova remnants, up to milligauss levels \citep{parizot06,reynolds21}, radio-emitting electrons have energies in the GeV range and radiative lifetimes against synchrotron losses of many thousands of years.  (Such electrons can also upscatter any ambient photon fields and lose energy due to such inverse-Compton scattering, but this will always be a subordinate loss mechanism unless radiation energy densities are comparable to that corresponding to $B \sim 100$ $\mu$G, or $u_{rad} \gapprox 10^{-9}$ erg cm$^{-3}$, much higher than generally found in or near SNRs.) 
The terms involving $b$ that result from the exponential cutoffs in the predicted equations for rates of flux change with time then contribute insignificantly in Equation~\ref{sxoft}, since at radio frequencies $\nu \ll \nu_p$
and $b \ll 1$.  The remaining behavior of flux with time is then
simple, depending only on the electron spectral index $s$ and expansion
index $m$ through $\beta$. 
As an example, for $s = 2.2$ (a typical value for other young SNRs as well) 
and $m = 0.7$, $\beta = 1.02$. For sufficiently strong deceleration 
($m < (s+5)/(s+11)$ or 0.55 for $s = 2.2$), $\beta < 0$ and the radio flux 
declines with time.  

\subsection{X-ray emission}

When $b \sim 1$, i.e., at X-ray wavelengths, 
measuring the spectral slope in X-rays constrains $\nu/\nu_p$.
Since 
\begin{equation}
    \left( \frac{\nu}{\nu_p} \right)^{1/2} = b \tau^\gamma
\end{equation}
we have, from Equation~\ref{slope}, 
%We can then just generalize Equation~\ref{xslopeV}: and 
\begin{equation}
    \alpha_x \equiv \Gamma - 1 = 
    \alpha + \frac{1}{2}\left( \frac{\nu}{\nu_p} \right)^{1/2}
    = \alpha + \frac{1}{2} b \tau^\gamma.
\label{xslopem}
\end{equation}
%where $\tau \equiv t/t_0$ as before.

\section{Discrete emission volumes}
\label{AppC}

Here we consider a fluid volume of relativistic electrons and magnetic field
which can be compressed or expanded without additional field or particles
generated.  A relativistic fluid of fast particles and field has an adiabatic
index of $4/3$ if radiative losses can be neglected.  (We shall assume our
discrete volume is dominated by nonrelativistic fluid, so that radiative losses
by the energetically unimportant relativistic electrons can be neglected in
the overall energetics, though of course they will affect the electron spectrum.)
So in a volume $V$, as above we have 
\begin{equation}    
     u_e \cong \frac{K}{s-2}E_l^{2-s} \propto V^{-4/3}
\label{ueV}    
\end{equation}
where again we take $s > 2$ (i.e., $\alpha > 0.5$)
%as is observed in all young SNRs, at least at radio-emitting energies)
and $E_h \gg E_l$.  
If the lower limit of the distribution $E_l$ is assumed constant in time, $K \propto V^{-4/3}.$  
%We shall make this assumption, though we discuss it further below.
Similarly,
\begin{equation}
    u_B = \frac{B^2}{8\pi} \propto V^{-4/3} \Rightarrow B \propto V^{-2/3}.
    \label{uBV}
\end{equation}
Here we have assumed that the magnetic field is disordered on dimensions smaller
than our emitting volume, which also means we assume the emission is isotropic.

Individual-particle energies vary as $V u_e \propto V^{-1/3}$, including the
$e$-folding energy $E_{\rm max}$ of any exponential cutoff.  A changing emission 
volume then changes the fiducial frequency $\nu_p$
characterizing the cutoff in the electron spectrum, due to changes in
both $E_{\rm max}$ and $B$:
\begin{equation}
    \nu_p \propto E_{\rm max}^2 B \propto V^{-4/3}.
\end{equation}

The flux from our discrete region then varies with volume in a complex way.
Define $j_0$ as the emissivity for some initial volume $V_0$, $\nu_{p0}$ the
maximum frequency at volume $V_0$, $b \equiv \sqrt{\nu/\nu_{p0}}$, and the 
fractional volume change $f \equiv V/V_0$.  Then $K/K_0 = f^{-4/3}$ for
constant $E_l$, etc., and 
\begin{equation}\sqrt{\nu/\nu_p} = f^{2/3}\sqrt{\nu/\nu_{p0}}\equiv f^{2/3}\, b.
\label{expfactor}
\end{equation}

Then we have, for constant $E_l$,
\begin{eqnarray}
    j_\nu  &=& c_j K B^{(s+1)/2} \nu^{-\alpha} 
      \exp{\left( -\sqrt{\nu/\nu_p}\right)}   \\
%    &=& c_j K_0 f^{-4/3} B_0^{(s+1)/2} f^{-(s + 1)/3} \nu^{-\alpha}
%    \exp{\left[ -\left( f^{2/3} \sqrt{\nu/\nu_{m0}}\right) \right]}\\
    &=& c_j K_0 B_0^{(s + 1)/2} \nu^{-\alpha}f^{-(s+5)/3}
       \exp{\left( {-bf^{2/3}}\right)}
\end{eqnarray}
since $b \equiv \sqrt{\nu/\nu_{p0}}$.  Then
\begin{equation}
    \frac{j_\nu}{j_\nu(V_0)} = f^{-(s+5)/3} \exp{\left( -bf^{2/3} -1 \right)}.
\end{equation}

So $S_\nu \equiv  4\pi \int j_\nu\,dV \cong 4\pi j_\nu\,V$
and
\begin{equation}
   \frac{S_\nu}{S_\nu(V_0)}  
   %& = & f^{-(s + 5)/3 + 1}
%     \exp{\left( -b( f^{2/3} - 1 )\right)} \\
   = f^{-(s+2)/3} \exp{\left( -b (f^{2/3} - 1) \right)}.
  \label{Vdep}
\end{equation}

If $E_l$ also evolves adiabatically, $E_l \propto V^{-1/3}$, adding additional $V$-dependence to Equation~\ref{Vdep}, through $K$:
\begin{equation}
K\,E_l^{2-s} \propto V^{-4/3} \ \ \ {\rm so} \ \ \ K \propto V^{-4/3 - (s-2)/3}
  = V^{-(s+2)/3}
  \end{equation}
and
\begin{equation}
\frac{S_\nu}{S_\nu(V_0)} 
%= f^{(1 - (s+2)/3 - (s+1)/3)} \exp{\left( -b (f^{2/3}-1) \right)}\\
= f^{-2s/3} \exp{\left( -b (f^{2/3} - 1)\right)}.
  \label{Vdep2}
\end{equation}
(We note that since $s$ is only slightly larger than 2, the volume-dependence of
the flux is almost the same in the two cases, Equations~\ref{Vdep} and~\ref{Vdep2}.)

The rate of change of fractional flux density $r \equiv S_\nu/S_\nu(V_0)$ with $f$ is then,
for constant $E_l$,
\begin{equation}
%%\ \Rightarrow \ 
%  \frac{d \ln S}{d \ln V} = 
  \frac{d \ln r}{d \ln f} = -\frac{s + 2}{3} 
    - \frac{2b}{3}f^{2/3}.
    \label{Vrate}
\end{equation}
  Similarly, for $E_l \propto V^{-1/3},$
\begin{equation}
    \frac{d \ln r}{d \ln f} = -\frac{2s}{3} - \frac{2b}{3}f^{2/3}.
    \label{Vrate2}
\end{equation}
Then time-dependence would come about through a factor $dV/dt$.
Take $V \propto t^z$, and define a reference time $t_0$ so that $V_0 \equiv V(t_0)$.  Then $f = (t/t_0)^z \equiv \tau^z$ and $d\ln r/d\ln \tau = z\,(d\ln r/d\ln f)$:
\begin{equation}
    \frac{d \ln r}{d \ln \tau} = \left( -\frac{s+2}{3} - \frac{2b}{3}\tau^{2z/3}\right) z \equiv q 
    \label{Vtimerate}
\end{equation}
for constant $E_l$, or \begin{equation}
q = \left( -\frac{2s}{3} - \frac{2b}{3}\tau^{2z/3}\right) z
\label{Vtimerate2}    
\end{equation}
for $E_l \propto V^{-1/3}$.  Then $S_\nu \propto t^q$.

\subsection{Radio emission} \label{radiomodel}

Again, $b \ll 1$ at radio frequencies, and the behavior of flux with volume 
change is then
simple, depending only on the electron spectral index, and obviously
accommodating either increases (for decreasing volume) or decreases (for
increasing volume, explaining the negative signs in Equations~\ref{Vrate} and~\ref{Vrate2}).  

\subsection{X-ray emission}

For typical X-ray observations, $b \sim 1$.  This can result in significant
additional contributions to the fractional flux rate of change, increasing
its absolute value (that is, there is greater leverage for changes in volume
altering the flux where the spectrum is steeper).  For compressions, $\ln f < 0$,
so $r$ increases.

As above, we can calculate the effect of a volume change on the spectral slope in X-rays:

\begin{equation}
\frac{d \ln S_\nu}{d \ln \nu} = 
  -\left[ \alpha + \frac{1}{2} \left( \frac{\nu}{\nu_p} \right)^{1/2} \right].
\label{xslope}
  \end{equation}

Again, measuring the spectral slope in X-rays constrains $\nu/\nu_p$. Using
Equation~\ref{expfactor} in the last equality,
\begin{equation}
    \alpha_x \equiv \Gamma - 1 = 
    \alpha + \frac{1}{2}\left( \frac{\nu}{\nu_p} \right)^{1/2}
    = \alpha + \frac{1}{2} b f^{2/3}.
\label{xslopeV}
\end{equation}

\bibliography{g1p9.bib}{}
\bibliographystyle{aasjournal}

\end{document}